\documentstyle[aps,prb,twocolumn,floats,psfig]{revtex}

\begin{document}

\bibliographystyle{prsty}

\title{
\begin{flushleft}
{\small \em submitted to}\\
{\small PHYSICAL REVIEW B \hfill $\qquad$ VOLUME
XX, NUMBER
XX
 \hfill
MONTH XX, YEAR XXXX }
\end{flushleft}
Dislocation-induced spin tunneling in Mn$_{12}$ acetate \vspace{-1mm} }

\author{
D. A. Garanin and E. M. Chudnovsky  }

\address{
Department of Physics and Astronomy, Lehman College, City University of New York, \\
250 Bedford Park Boulevard West, Bronx, New York 10468-1589 \\
\smallskip
{\rm(Received 27 May 2001)}
\bigskip\\
\parbox{14.2cm}
{\rm Comprehensive theory of quantum spin relaxation in Mn$_{12}$ acetate crystals is
developed, that takes into account imperfections of the crystal structure and is based upon
the generalization of the Landau-Zener effect for incoherent tunneling from excited energy
levels. It is shown that linear dislocations at plausible concentrations provide the
transverse anisotropy which is the main source of tunneling in Mn$_{12}$. Local rotations of
the easy axis due to dislocations result in a transverse magnetic field generated by the field
applied along the $c$-axis of the crystal, which explains the presence of odd tunneling
resonances. Long-range deformations due to dislocations produce a broad distribution of tunnel
splittings. The theory predicts that at subkelvin temperatures the relaxation curves for
different tunneling resonances can be scaled onto a single master curve. The magnetic
relaxation in the thermally activated regime follows the stretched-exponential law with the
exponent depending on the field, temperature, and concentration of defects.
\smallskip
\begin{flushleft}
\end{flushleft}
} } \maketitle

\section{Introduction}
\label{sec_intro}

Physical properties of Mn$_{12}$ acetate crystals have been subject of intensive investigation
in the last years.
The crystals have been chemically synthesized by Lis\cite{lis80} in 1980, who found that they
had a centered tetragonal structure with $a=17.319 \AA$ and $c=12.388 \AA$ as lattice
parameters. \cite{lis80,henetal97}
At the beginning of 1990's Sessoli {\em et al} of the Florence group of chemists (see, e.g.,
Refs.\ \onlinecite{sesgatcannov93,ses95,novses95}) established that Mn$_{12}$ clusters located
at the sites of the crystal lattice have spin 10.
Sessoli {\em et al} also demonstrated that Mn$_{12}$ crystals are characterized by a record
high uniaxial magnetic anisotropy along the $c$-direction, that makes the 65K energy barrier
between the spin-up and spin-down states.
The explosion of the interest of physicists to Mn$_{12}$ acetate came after Friedman {\em et
al}$\,$ \cite{frisartejzio96} measured its spectacular stepwise magnetic hysteresis and
explained it by resonant spin tunneling (see also the followup experiments, Refs.\
\onlinecite{heretal96,thoetal96}).
Numerous experiments on Mn$_{12}$ performed since 1996 uncovered a number of other interesting
phenomena, such as memory effects, \cite{wersesgat99} non-exponential relaxation,
\cite{wersesgat99,peretal98,bokkenwal00} and a peculiar crossover between thermal and quantum
behavior \cite{kenetal00,bokkenwal00} predicted by theory.
\cite{garchu97,chugar97,garmarchu98,liamueparzim98,leemueparzim98,%
zhaetal99,kim99,garchu99,garchu00,paryooyoo00,chokim00,marchu00}

\begin{figure}[t]
\unitlength1cm
\begin{picture}(11,9)
\centerline{\psfig{file=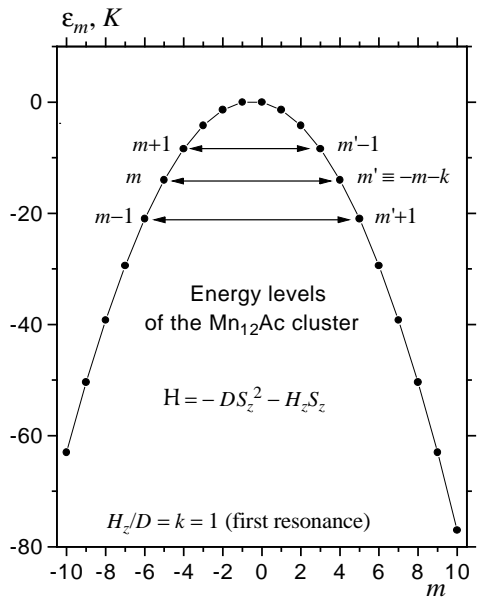,angle=0,width=8cm}}
\end{picture}
\caption{ \label{fig_levels} Spin energy levels of a Mn$_{12}$ Ac molecule for $H_x=0$ and
$H_z = D$ corresponding to the first resonance, $k=1$. }
\end{figure}

Uniaxial spin Hamiltonian, in the field parallel to the anisotropy axis, ${\cal H} = -DS_z^2
-H_zS_z$, has pairs of degenerate levels for
\begin{equation}\label{ResCond}
H_z=kD, \qquad k=0,\pm 1, \pm 2,\ldots, \pm(2S-1).
\end{equation}
For a sample initially magnetized in the negative $z$-direction, Mn$_{12}$ molecules occupy
spin states with a negative magnetic quantum number $m$.
When the magnetic field is applied in the positive $z$-direction, the molecules eventually
relax to the states with positive $m$.
It may occur via thermal activation over the anisotropy barrier or through quantum tunneling
between the states on different sides of the barrier (see Fig.\ \ref{fig_levels}).
The latter process adds to the thermal activation when the levels on the two sides of the
barrier are on resonance, that is, at $H_z=kD$.
Consequently, at $H_z=kD$ the magnetic relaxation of the crystal is faster than for the
off-resonance values of the field. \cite{frisartejzio96}

There exist two macroscopic experimental approaches to the study of spin tunneling in
Mn$_{12}$ (see, e.g., Refs.\
\onlinecite{frisartejzio96,heretal96,thoetal96,frietal97,luisetal97,%
frisarzio98,foretal98,gometal98,peretal98,wersesgat99,thocanbar99,%
bokkenwal00,kenetal00,zhosaryoohen00,chietal00} and references therein).
Both clearly demonstrate the effect of resonant spin tunneling.
In the first approach, one measures the magnetic relaxation in the crystal at fixed
temperature and magnetic field.
The theory of such a relaxation works out tunneling transitions between levels on two sides of
the energy barrier and spin-phonon transitions between the levels on one side of the barrier.
These processes are described by the density-matrix formalism suggested in Ref.\
\onlinecite{garchu97} and further developed in Refs.\
\onlinecite{vilwueforret97,luisetal97,leulos99epl,leulos00prb}.
In the second approach, one sweeps magnetic field through the resonant value and measures the
fraction of molecules that change their magnetic moments.
For the ground-state tunneling, the theory of such process has been developed along the lines
of the Landau-Zener effect. \cite{lan32,zen32,dobzve97,gun97,leulos00zen,garchu01prl}
A nice feature of the field-sweep approach is that the fraction of molecules that change the
direction of the spin depends on the tunneling level splitting and on the sweep rate but is
insensitive to the dissipation in the experimental limit of small splitting.

While the uniaxial Hamiltonian allows to compute resonant fields from independent measurements
of $D$, it does not explain why spin tunneling in Mn$_{12}$ actually occurs.
This is because ${\cal H} = -DS_z^2 -H_zS_z$ conserves $S_z$ and thus does not allow any
transitions between different $m$.
To obtain tunneling, one should include in the Hamiltonian the terms which do not commute with
$S_z$.
The tetragonal symmetry of the crystal does not allow transverse anisotropy terms which are
quadratic on the spin operator, i.e., terms proportional to $S_x^2$ and $S_y^2$.
The lowest order transverse terms must be proportional to $S_+^4 + S_-^4$. \cite{harpolvil96}
If these were the only terms in the Hamiltonian responsible for tunneling, then only
resonances with ``tunneling length'' $m'-m$ (see Fig.\ \ref{fig_levels}), which is a multiple
of four, would have been observed.
This is not the case for Mn$_{12}$.
To explain the presence of resonances with all $k$, one should invoke a transverse field
contribution to the Hamiltonian, $-H_x S_z$.
Dipolar fields from magnetic moments of Mn$_{12}$ molecules and hyperfine fields from Mn
nuclei have been suggested as natural candidates.
\cite{harpolvil96,garchu97,prosta98,garchusch00,chu00prl,prosta00,%
werpauses00}
Experiments, however, indicate that transverse fields needed to explain the data are stronger
than the ones produced by atomic nuclei and magnetic dipoles.
The nature of the effects responsible for tunneling in Mn$_{12}$ remains an open question.

Next controversy is the origin of the $\sqrt{t}$ law that approximately fits the initial stage
of the relaxation in some experiments. \cite{wersesgat99,bokkenwal00,chietal00}
Suggested explanations include collective effects due to dipolar interaction between Mn$_{12}$
molecules \cite{prosta98} and fluctuating random noise. \cite{miysai00}
The first requires some special initial conditions which are not satisfied in experiment
\cite{chu00prl,prosta00,werpauses00} while the second requires certain assumptions about the
spectrum of fluctuations.

Another open question is the nature of the ``minor species'' of Mn$_{12}$ that relaxes faster
than the ``major species'', and is present in all samples studied to date.
In the subkelvin temperature range, when the relaxation slows down, the minor species is the
main source of the relaxation seen in experiment.
Wernsdorfer {\em et al} attribute that species to the defective sites of the crystal lattice
(see Footnote 9 in Ref.\ \onlinecite{wersesgat99}).
On general grounds, it is obvious that defects inside the Mn$_{12}$ clusters cannot account
for the minor species.
Such defects would strongly modify the spin Hamiltonian, which would result in a significant
change of the resonant fields and orders-of-magnitude change of tunneling rates, in contrast
with experimental observations.

On the contrary, defects of the tetragonal crystal lattice such as dislocations, which do not
change the structure of the cluster, should produce symmetry-violating terms in the
Hamiltonian that are responsible for tunneling.
Some of these terms are {\em quadratic} in $S_x$ and $S_y$.
For even $k$ [see Eq.\ (\ref{ResCond})] they produce tunneling in a lower order of the
perturbation theory than the transverse-field.
In addition, dislocations give rise to local rotations of the easy axis, which for $H_z\neq 0$
results in a {\em transverse field} that unfreezes resonances with odd $k$.

Because of the long-range nature of deformations caused by dislocations, the number of
affected crystal sites should be relatively large even for a moderate concentration of
dislocations.
Strong deformations will exist only at a small number of sites inside dislocation cores.
Most of the Mn$_{12}$ molecules will develop weak deformations, with the tunneling rate
depending on the location of the molecule.
Consequently, the relaxation process evolves from the relaxation of the minor species, close
to the dislocation cores, to the relaxation of the major species far from the dislocation
cores.

\begin{figure}[t]
\unitlength1cm
\begin{picture}(11,5.3)
\centerline{\psfig{file=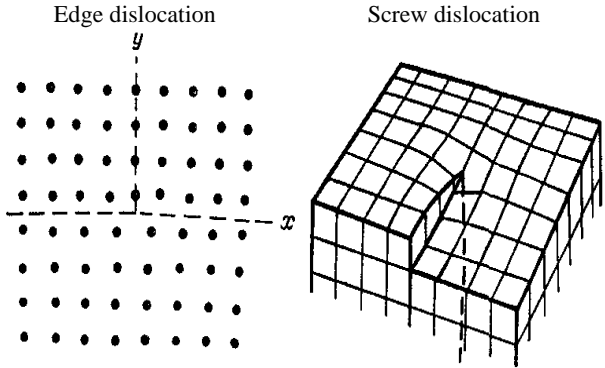,angle=-90,width=9cm}}
\end{picture}
\caption{ \label{fig_edgescr} Edge and screw dislocations. }
\end{figure}

In this paper we show that edge and screw dislocations (see Fig.\ \ref{fig_edgescr}) should be
the main source of spin tunneling in Mn$_{12}$ crystals.
Broad distribution of deformations causes broad distribution of thermal activation and
tunneling rates.
To study the magnetic relaxation in Mn$_{12}$, we develop the theory of incoherent
Landau-Zener tunneling from excited energy levels.
We compute the relaxation law in a crystal with dislocations and show that it obeys scaling
that can be tested in experiment.
We also demonstrate that in the kelvin temperature range the relaxation is well described by
the stretched-exponential law.

The paper is organized as follows.
Spin-lattice couplings due to different types of dislocations are studied in Sec.\
\ref{sec_couplings}.
Sec.\ \ref{sec_splittings} contains perturbation formulas for the tunneling level splittings
for even and odd resonances.
The  analytical formula for the distribution of transverse anisotropies due to a random array
of dislocations is derived in Sec.\ \ref{sec_random}.
The distribution of tunnel splittings in Mn$_{12}$ crystals is studied in Sec.\
\ref{sec_distrsplittings}.
Sec.\ \ref{sec_landauzener} is devoted to the incoherent Landau-Zener transitions at finite
temperatures.
In Sec.\ \ref{sec_sweeping} the results are applied  to Mn$_{12}$ crystals with the
distribution of tunel splittings.
In Sec.\ \ref{sec_activated} we compute the magnetic relaxation in the thermal activation
regime.
The implications of the results obtained in this paper are summarized in Sec.\
\ref{sec_conclusion}.

%

\section{Spin-lattice couplings due to dislocations}
\label{sec_couplings}

\subsection{Spin Hamiltonian}

We study the Hamiltonian
\begin{equation}\label{ham}
{\cal H} = - D S_z^2 - H_z S_z + {\cal H}_{\rm me},
\end{equation}
where $S_z$ is the $z$-component of the spin operator, $S=10$, $D=0.65$~K, $H_z$ is the
magnetic field applied along the $z$-axis ($c$-axis of the crystal), and ${\cal H}_{\rm me}$
is the magnetoelastic coupling.
The Hamiltonian of Mn$_{12}$  also contains crystal fields of fourth order on the spin
operator, magnetic dipole interactions, and hyperfine interactions.
We neglect them in order to emphasize the effect of dislocations.
The magnetoelastic coupling in Mn$_{12}$ is of the form
\cite{harpolvil96,leulos99epl,leulos00prb}
\begin{eqnarray}\label{hamme}
&& {\cal H}_{\rm me} = g_1 D ( \varepsilon_{xx} - \varepsilon_{yy} ) (S_x^2-S_y^2) + g_2 D
\varepsilon_{xy} \{ S_x, S_y \}
\nonumber\\
&& \qquad {} + D ( \alpha_x \{ S_x, S_z \} + \alpha_y \{ S_y, S_z \} ),
\end{eqnarray}
where $\{ \hat A, \hat B \}$ is anticommutator,
\begin{equation}\label{alphaDef}
\alpha_x \equiv g_3 \varepsilon_{xz} + g_4 \omega_{xz}, \qquad \alpha_y \equiv g_3
\varepsilon_{yz} + g_4 \omega_{yz} ,
\end{equation}
 and
\begin{equation}\label{epsDef}
\epsilon_{{\alpha}{\beta}}=\frac{1}{2}\left(\frac{\partial u_\alpha} {\partial x_ \beta} +
\frac{
\partial u_ \beta }{\partial x_\alpha} \right), \quad \omega_{ \alpha \beta }
=\frac{1}{2}\left(\frac{ \partial u_ \alpha } { \partial x_ \beta } - \frac{ \partial u_ \beta
}{
\partial x_ \alpha }  \right)
\end{equation}
are linear deformation tensors; $\omega_{ \alpha \beta }$ being a pure rotation and ${\bf u}$
being the displacement.
Note that, in principle, the Hamiltonian may contain terms of higher order on deformations.
The exact strain tensor also has quadratic contribution on displacement.
These terms may become important at and near dislocation cores, that is, for a small number of
Mn$_{12}$ molecules which do not contribute much to the relaxation process.

The effect of dislocations is determined by values of constants $g_1$, $g_2$,$g_3$, and $g_4$.
The value of $g_4$ is fixed by the symmetry of the crystal, $g_4=1$.
\cite{harpolvil96,garchu97,chugar00epl}
There are general arguments that all $g$-constants must be of the same order, see Ref.\
\onlinecite{leulos00epl} and references therein.
For numerical work we will use  $g_1=g_2=g_3=g_4=1$.

Before calculating tunneling rates of Mn$_{12}$ molecules due to dislocations, it is
convenient to make rotations of the spin axes in $(x,y)$, $(x,z)$, and $(y,z)$ planes in order
to transform the Hamiltonian of Eq.\ (\ref{ham}) into a diagonal quadratic form on spin
operators.
The terms $\{ S_x, S_z \}$ and $\{ S_y, S_y \}$ in Eq.\ (\ref{hamme}) are eliminated by
rotations by small angles, $\alpha_x/2$ and $\alpha_y/2$, in the $(x,z)$ and $(y,z)$ planes,
respectively.
This results in a slight renormalization of the uniaxial anisotropy constant $D$ and of the
transverse-anisotropy terms.
Both effects are quadratic on deformations and are neglected in
our approach.
The rotations also change the form of the Zeeman term $-H_zS_z$, resulting in the {\em
transverse} field
\begin{equation}\label{TransvField}
H_\perp \cong \frac 12 \sqrt{\alpha_x^2 + \alpha_y^2} H_z
\end{equation}
in the rotated coordinate system, which is linear on deformations and proportional to the
magnetic field $H_z$ applied along the $c$-axis of the crystal.
This transverse field unfreezes tunneling transitions which change the spin projection $m$ by
an odd number.
The terms $\{ S_x, S_y \}$ in Eq.\ (\ref{hamme}) are eliminated by rotations in the $(x,y)$
plane by an angle which is not small.
They produce the transverse anisotropy $E(S_{x'}^2-S_{y'}^2)$ in the first order on
deformations.

In the four subsections below, we will consider different types of linear dislocations. We
will use formulas of isotropic elastic theory for displacements, \cite{lanlif7} which are
accurate enough for our conclusions and, at the same time, are less cumbersome than the exact
expressions for the tetragonal crystal symmetry.
We will see that for all types of linear dislocations the spatial dependence of the transverse
anisotropy and the transverse field can be written as
\begin{equation}\label{EHtrGen}
E = 2D\frac {g(\varphi)}{r}, \qquad H_\perp = H_z \frac {g_H(\varphi)}{r},
\end{equation}
where $r$ is the distance from the dislocation axis, measured in the lattice units, whereas
$g(\varphi)$ and $g_H(\varphi)$ are functions of the angle, which are of order one if $g_1
\sim g_2 \sim g_3 \sim g_4 =1$.
One can immediately see from Eq.\ (\ref{EHtrGen}) that the effect of dislocations on tunneling
must be strong.
Indeed, for $r\sim 1$ one has $E\sim D$, whereas the spatial decay of $E$ is slow, so that
each dislocation affects a large number of Mn$_{12}$ molecules in the crystal.

\subsection{Screw dislocation along the anisotropy axis}

For screw dislocations shown in Fig.\ \ref{fig_edgescr} the only nonzero component of the
displacement ${\bf u}$ is $u_z$.
For the dislocation axis at $x,y=0$ it is given by
\begin{equation}\label{DisplScrewZ}
u_z(x,y) = \frac c {2\pi} \arctan \frac y x,
\end{equation}
where $c$ is the lattice spacing along the $c$-axis.
The only nonzero components of deformations are
\begin{eqnarray}\label{DefScrewZ}
&& %
\varepsilon_{xz} =-\omega_{xz} = - \frac c {4\pi} \frac y {x^2 +
y^2} \nonumber\\
&& %
\varepsilon_{yz} =-\omega_{yz} = \frac c {4\pi} \frac x {x^2 + y^2}
\end{eqnarray}
Thus, according to Eq.\ (\ref{hamme}), screw dislocations along the $c$-axis do not produce
transverse anisotropy of the first order in deformations.
They, nevertheless, create a transverse field in the rotated coordinate system, which is given
by
\begin{equation}\label{TransvFieldScrewZ}
H_\perp \cong \frac c {8\pi} \frac {|g_3-g_4|} {\sqrt{x^2 + y^2}} H_z
\end{equation}
[cf.\ Eq.\ (\ref{TransvField})].
This transverse field gives rise to spin tunneling of order $H_\perp^{m'-m}$ between the
resonant pair of levels $\varepsilon_m$ and $\varepsilon_{m'}$ [see Eq.\ (2.6) of Ref.\
\onlinecite{garchu97}].
Since $m'-m$ is a usually a large number, this tunneling is much weaker than tunneling due to
the transverse anisotropy $E$ from other types of dislocations, which appears in the
$E^{(m'-m)/2}$ order for even resonances (see Sec.\ \ref{sec_splittings} for more detail).
Since all types of dislocations must be simultaneously present in a crystal, we conclude that
$c$-axis screw dislocations are irrelevant for spin tunneling.

\subsection{Screw dislocation perpendicular to the anisotropy axis}

For a screw dislocation along the $Y$-axis the only nonzero component of the displacement is
\begin{equation}\label{DisplScrewY}
u_y(x,z) = \frac b {2\pi} \arctan \frac z x,
\end{equation}
where $b$ is the appropriate lattice spacing.
Thus the nonzero components of the deformation tensors are
\begin{eqnarray}\label{DefScrewY}
&& %
\varepsilon_{yz} =\omega_{yz} = \frac b {4\pi} \frac x {x^2 +
z^2} \nonumber\\
&& %
\varepsilon_{xy} = - \frac b {4\pi} \frac z {x^2 + z^2} .
\end{eqnarray}
To bring the spin Hamiltonian to the canonical form, one should first rotate the spin axes in
the ($y,z$) plane by an angle $\alpha_y/2$, which results in a transverse field
\begin{equation}\label{TransvFieldScrewY}
H_\perp \cong \frac 12 \alpha_y H_z = (g_3+g_4) \frac b {8\pi} \frac {x} {x^2 + y^2} H_z
\end{equation}
directed along the new $Y$-axis.
Then the rotation by $\phi=\pi/4$ in the $x,y$ plane transforms $\{S_x,S_y\}$ into $-(S_{x'}^2
- S_{y'}^2)$, which results in the Hamiltonian
\begin{equation}\label{HamRotScrewY}
{\cal H} = - D S_{z'}^2 - H_z S_{z'} + E (S_{x'}^2 - S_{y'}^2)  - H_{x'} S_{x'} - H_{y'}
S_{y'},
\end{equation}
where
\begin{equation}\label{EScrewY}
E = -g_2 D \varepsilon_{xy} = g_2 D \frac b {4\pi} \frac z {x^2 + z^2}
\end{equation}
and
\begin{equation}\label{HtransvScrewY}
H_{x'} = H_{y'} = (g_3+g_4) \frac b {8\sqrt{2}\pi} \frac {x} {x^2 + y^2} H_z
\end{equation}
(the signs of $H_{x'}$ and $H_{y'}$ are irrelevant).
This type of dislocations generates the spatially-dependent transverse anisotropy which is the
main source of tunneling.
It also generates a transverse field that unfreezes odd resonances.
The hard and medium axes, $x'$ and $y'$, interchange when one crosses the plane $z=0$, whereas
the transverse field always has equal components along the hard and medium axes.

\begin{figure}[t]
\unitlength1cm
\begin{picture}(11,6.2)
\centerline{\psfig{file=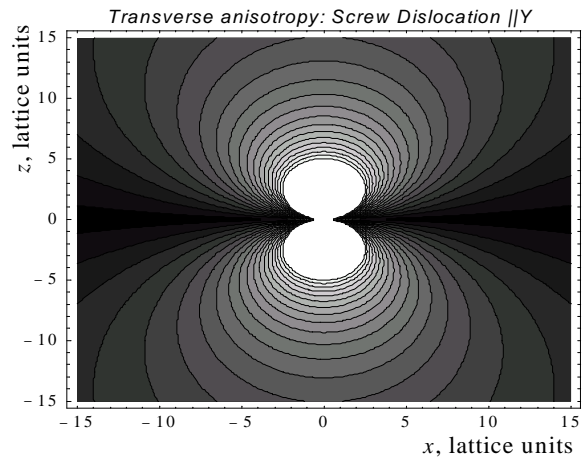,angle=-90,width=9cm}}
\end{picture}
\caption{ \label{fig_escrew} Contour plot of transverse anisotropy $E$ created by a screw
dislocation along the $Y$-axis, see Eq.\ (\protect\ref{EScrewY}).
(Grey scales are arbitrary.) }
\end{figure}
\begin{figure}[t]
\unitlength1cm
\begin{picture}(11,6.5)
\centerline{\psfig{file=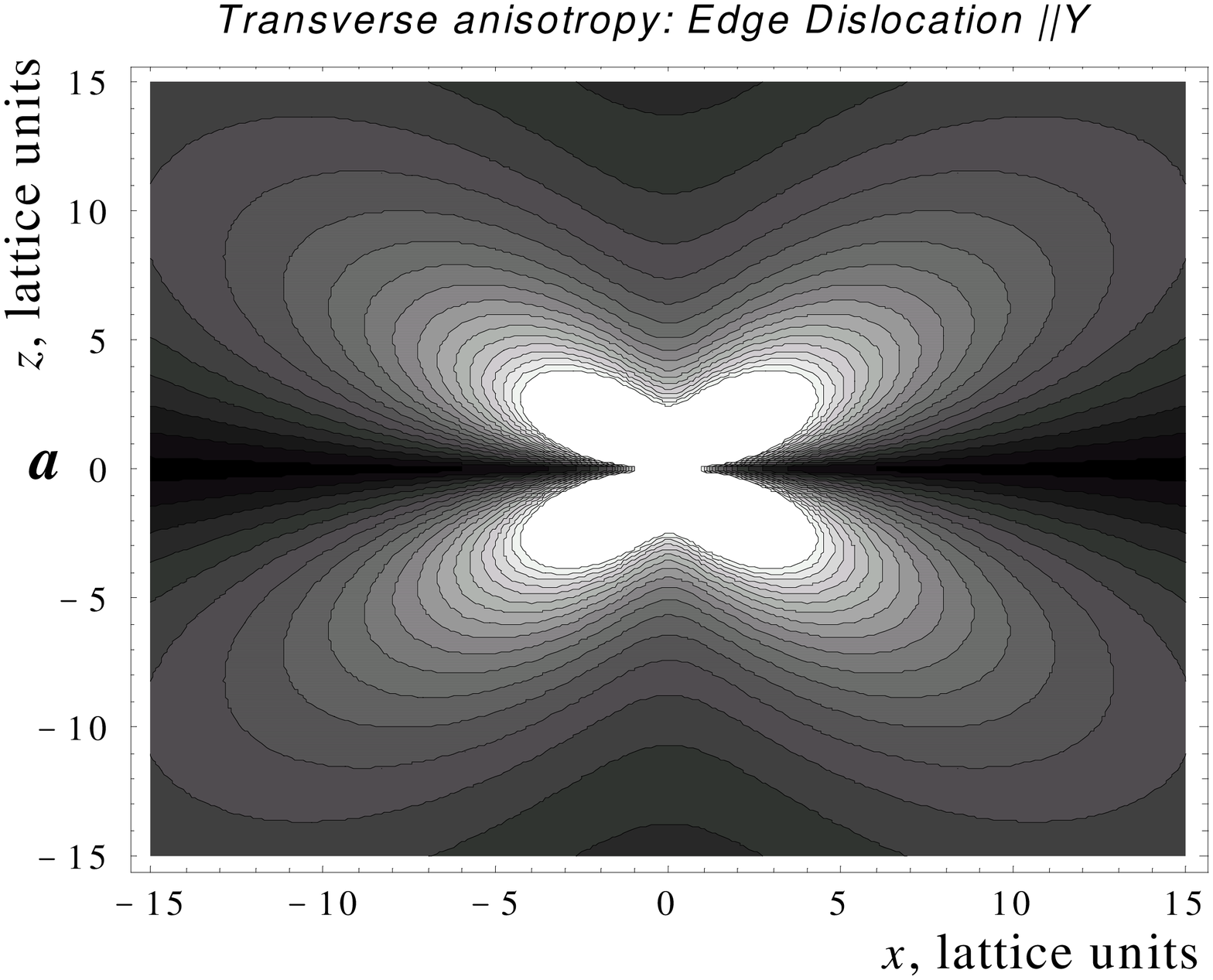,angle=-90,width=9cm}}
\end{picture}
\begin{picture}(11,8.0)
\centerline{\psfig{file=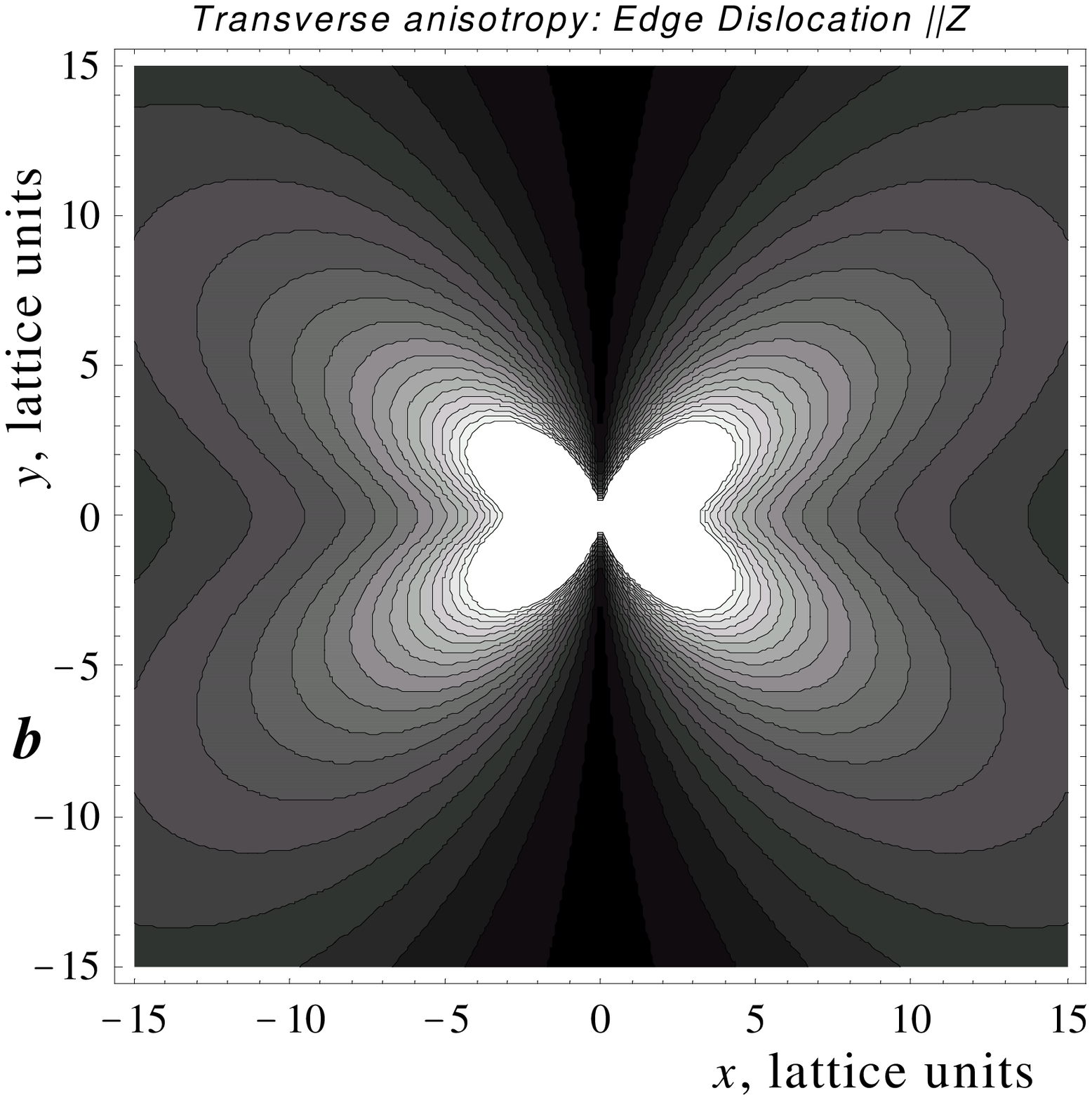,angle=-90,width=11.7cm}}
\end{picture}
\caption{ \label{fig_eedge} Contour plots of transverse anisotropy $E$ created by: $a$ - an
edge dislocation along the $Y$-axis, see Eq.\ (\protect\ref{EHxDef}); $b$ - an edge
dislocation along the $Z$-axis, see Eq.\ (\protect\ref{EEdgeZ}).
 }
\end{figure}
%

\subsection{Edge dislocation along the anisotropy axis}

For this dislocation shown in Fig.\ \ref{fig_edgescr} there is no displacement along the $c$
axis, i.e., $u_z=0$, whereas other displacement components are given by
\begin{equation}\label{xDisplEdgeZ}
u_x = \frac b {2\pi} \left[ \arctan\frac y x + \frac 1 {2(1-\sigma)} \frac {xy} {x^2 + y^2}
\right]
\end{equation}
and
\begin{equation}\label{yDisplEdgeZ}
u_y = -\frac b {2\pi} \left[ \frac{1-2\sigma}{4(1-\sigma)} \ln(x^2 + y^2) + \frac 1
{2(1-\sigma)} \frac {x^2} {x^2 + y^2} \right],
\end{equation}
where $0<\sigma <1/2$ is the Poisson elastic coefficient (we will use $\sigma=0.25$ in the
numerical work).
For the deformations one obtains
\begin{equation}\label{xxyyDeflEdgeZ}
\varepsilon_{xx}-\varepsilon_{yy} = -\frac b {\pi (1-\sigma)}\frac {x^2 y} {(x^2 + y^2)^2}
\end{equation}
and
\begin{equation}\label{xyDeflEdgeZ}
\varepsilon_{xy} = -\frac b {4\pi (1-\sigma)} \frac {x(x^2-y^2)} {(x^2 + y^2)^2}.
\end{equation}
It is clear that this dislocation does not tilt the anisotropy axis and thus does not generate
a transverse field.
After the rotation in the ($x,y$) plane the spin Hamiltonian becomes
\begin{equation}\label{HamRotEdgeZ}
{\cal H} = - D S_{z}^2 - H_z S_{z} + E (S_{x'}^2 - S_{y'}^2)
\end{equation}
with
\begin{equation}\label{EEdgeZ}
E = D \sqrt{g_1^2 (\varepsilon_{xx}-\varepsilon_{yy})^2 + g_2^2 \varepsilon_{xy}^2 }.
\end{equation}
%

\subsection{Edge dislocation perpendicular to the anisotropy axis}

For such dislocations the nonzero displacements are $u_x$ and $u_z$.
They are given by Eqs.\ (\ref{xDisplEdgeZ}) and (\ref{yDisplEdgeZ}) with $y \Rightarrow z$.
For deformations one obtains $\varepsilon_{xy}=0$,
\begin{equation}\label{xxyyDefEdgeY}
\varepsilon_{xx}-\varepsilon_{yy} = -\frac b {4\pi (1-\sigma)} z \frac{ (3-2\sigma)x^2 +
(1-2\sigma)z^2 } {(x^2 + z^2)^2}
\end{equation}
and
\begin{equation}\label{alxDefEdgeY}
\alpha_x = \frac 12 \left[ (g_3+g_4) \frac { \partial u_x } {\partial z} + (g_3-g_4) \frac {
\partial u_z } {\partial x} \right],
\end{equation}
where
\begin{equation}\label{uxzEdgeY}
\frac { \partial u_x } {\partial z} = \frac b {4\pi (1-\sigma)} x \frac{ (3-2\sigma)x^2 +
(1-2\sigma)z^2 } {(x^2 + z^2)^2}
\end{equation}
and
\begin{equation}\label{uzxEdgeY}
\frac { \partial u_z } {\partial x} = \frac b {4\pi (1-\sigma)} x \frac{ (1-2\sigma)x^2 +
(3-2\sigma)z^2 } {(x^2 + z^2)^2}.
\end{equation}
The resulting spin Hamiltonian has the form
\begin{equation}\label{HamRotEdgeY}
{\cal H} = - D S_{z'}^2 - H_z S_{z'} + E  (S_{x'}^2 - S_{y'}^2)  - H_{x'} S_{x'} ,
\end{equation}
where
\begin{equation}\label{EHxDef}
E = g_1 D ( \varepsilon_{xx} - \varepsilon_{yy} ), \quad H_{x'} = \frac 12 \alpha_x H_z.
\end{equation}
Above the XY plane (at $z>0$) the transverse field is directed along the hard axis, while at
$z<0$ the transverse field is along the medium axis.

\section{Tunnel splittings}
\label{sec_splittings}

For concentrations of dislocations upto $c=10^{-2}$ (in lattice units) most of Mn$_{12}$
molecules are still far from the dislocation cores.
Thus the transverse anisotropy and the transverse field generated by dislocations [see, e.g.,
Eq.\ (\ref{HamRotEdgeY})] are typically small in comparison with the uniaxial term in the
Hamiltonian of the ideal crystal.
In this situation one can obtain the tunneling splittings $\Delta_{mm'}$ of resonant pairs of
levels in the lowest-order of the perturbation theory.
For the model with the transverse field the splittings are \cite{garchu97,chufri,gar91jpa}
\begin{eqnarray}\label{splitH}
&& \Delta_{mm'} = \frac{2D}{[(m'-m-1)!]^2}
\nonumber\\
&& \qquad \times \sqrt{\frac{ (S+m')! (S-m)! }{ (S-m')! (S+m)! } } \left( \frac{ H_x }{ 2D }
\right)^{m'-m}.
\end{eqnarray}
For the transverse anisotropy model, the splittings have been calculated in Ref.\
\onlinecite{gar91jpa} for $H_z=0$ and generalized in Ref.\ \onlinecite{garchu99} for any field
bias.
It is convenient to rewrite Eq.\ (4) of Ref.\ \onlinecite{garchu99} in the form
\begin{eqnarray}\label{splitE}
&& \Delta_{mm'} = \frac{2D}{[(m'-m-2)!!]^2}
\nonumber\\
&& \qquad \times \sqrt{\frac{ (S+m')! (S-m)! }{ (S-m')! (S+m)! } } \left( \frac{ |E| }{ 2D }
\right)^{(m'-m)/2}.
\end{eqnarray}
In the transverse-anisotropy model, $\Delta_{mm'}$ is nonzero for even $m'-m = -2m-k$, which
for even S requires an even value of the tunneling resonance number $k$.
For odd values of $m'-m$, tunneling resonances are quenched, $\Delta_{mm'}=0$.

In Mn$_{12}$ samples with dislocations, both the transverse field and the transverse
anisotropy are present, see Eq.\ (\ref{EHtrGen}).
Since at resonances the longitudinal field satisfies $H_z=kD$, the transverse field is well
defined for a given type and configuration of dislocations, with $E$ and $H_\perp$ being of
the same order of magnitude.
To obtain tunnel splittings perturbatively in such a model, one has to sum up the products of
matrix elements of perturbations and corresponding energy denominators, each product
representing a chain connecting the states $m$ and $m'$. \cite{leulos00prb}
In general, this algorithm is very cumbersome and has no advantages with respect to direct
numerical diagonalization of the spin Hamiltonian.

\begin{figure}[t]
\unitlength1cm
\begin{picture}(11,6.5)
\centerline{\psfig{file=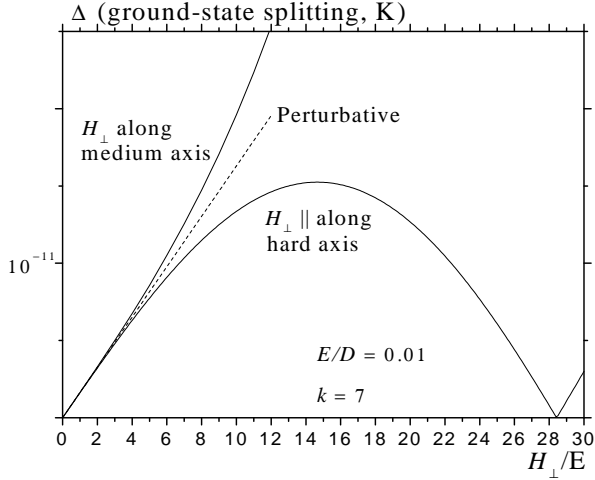,angle=-90,width=9cm}}
\end{picture}
\caption{ \label{fig_splvshx} Exact and perturbative results for the transverse-field
dependence of the splitting of the metastable ground-state at the $k=7$ tunneling resonance.}
\end{figure}

Fortunately, as will become apparent below, for $H_\perp \sim E \ll D$ the effect of the
transverse field on tunneling is much weaker than that of the transverse anisotropy.
Thus, for the even resonances one can neglect the field contribution to the splittings,
whereas for the odd resonances the field should be taken into account in the first order only.
In the latter case, the main source of the splitting is the transverse anisotropy taken in a
high order of a perturbation theory needed to build a chain of matrix elements that changes
the spin projection by $m'-m-1$.
For odd resonances the transverse field makes the missing single perturbation step along that
chain, needed to change the spin projection by one.
The corresponding matrix element $\langle m_1 | H_{x}S_{x} | m_1+1 \rangle$ can be inserted at
any place in the chain, $m_1=m, m+1,\ldots, m'-1$.
Quite fortunately, the corresponding sum can be calculated exactly:
\begin{equation}\label{GammaSum}
\frac 1 \pi \sum_{p=0}^r \frac { \Gamma(r-p+1/2)  \Gamma(p+1/2) } {\Gamma(r-p+1)  \Gamma(p+1)
} =1
\end{equation}
and the resulting splittings for odd resonances acquires the elegant form:
\begin{eqnarray}\label{splitEHx}
&& \Delta_{mm'} = \frac{H_x}{[(m'-m-2)!!]^2}
\nonumber\\
&& \qquad \times \sqrt{\frac{ (S+m')! (S-m)! }{ (S-m')! (S+m)! } } \left( \frac{ |E| }{ 2D }
\right)^{(m'-m-1)/2}.
\end{eqnarray}

Note that this perturbative result does not depend on the sign of $E$, that is, the splitting
is the same for the transverse field directed along the medium or hard axes.
The difference between these two cases appears in higher orders in the transverse field, see
Fig.\ \ref{fig_splvshx}.
For the transverse field along the hard axis, there is a destructive interference of the
contributions of the transverse field and the transverse anisotropy \cite{garg93} that cause
splittings to vanish at \cite{kecgar01,garchudiabolic}
\begin{equation}\label{SplQuench}
H_x =  k_x \sqrt{2E(D+E)}.
\end{equation}
For the integer spin $S$ the number $k_x$  satisfying the condition $|k_x|+|k|\leq 2S+1$ is
odd, where $k$ is the tunneling resonance number of Eq.\ (\ref{ResCond}).
For even resonances $k$ splittings vanish at $k_x=0,\pm 2, \ldots$, whereas for odd resonances
the splittings vanish at $k_x=\pm 1,\pm 3, \ldots$
It is clear from Eq.\ (\ref{SplQuench}) and from Fig.\ \ref{fig_splvshx} that the linear
dependence of the splitting on the transverse field, as given by Eq.\ (\ref{splitEHx}), is
valid for $H_\perp \ll  \sqrt{2ED}$, which is satisfied in our case $H_\perp \sim E \ll D$.

The condition on the transverse field obtained above also follows from simple perturbative
arguments.
For the matrix element of the perturbation $V_{m_1,m_1+2} = \langle m_1 | \hat V | m_1+2
\rangle$ one has $V_{m_1,m_1+2} \sim E$ if the perturbation is due to the transverse
anisotropy, and $V_{m_1,m_1+2} \sim H_\perp^2/D$ if it is due to the transverse field.
Thus the effects of both types of perturbation are comparable if $H_\perp \sim  \sqrt{ED}$.

Let us compare now the strengths of even and odd tunneling resonances.
For an odd resonance the ratio of $\Delta_{mm'}$ to the geometrical mean value of the
splittings of adjacent even resonances has the form
\begin{equation}\label{SplRatio}
\frac{\Delta_{mm'} }{ \sqrt{ \Delta_{m,m'+1} \Delta_{m,m'-1} }} = F(S,m,m') \frac {H_x}
{\sqrt{2ED}},
\end{equation}
where
\begin{eqnarray}\label{FRatio}
&&
 F(S,m,m') = \frac { (m'-m-1)!! (m'-m-3)!! }{ [(m'-m-2)!!]^2 }
\nonumber\\
&& \qquad {}\times \left[\frac {(S+m')(S-m'+1)}{(S-m')(S+m'+1)} \right]^{1/4}.
\end{eqnarray}
The function $F(S,m,m')$ is of order unity.
It approaches $\pi/2$ in the quasiclassical case of large $S$, $m$, and $m'$.
Thus the odd resonances are weaker than the even ones by the factor of order $H_x/\sqrt{DE}$.
This effect is not dramatic since odd and even resonances only differ by a prefactor in front
of high power of the small ratio $E/D$, see Eqs.\ (\ref{splitE}) and (\ref{splitEHx}).

It should be stressed that for numerical calculation of tunnel splittings the {\em exact}
resonance condition \cite{kecgar01,garchudiabolic}
\begin{equation}\label{ResCondExact}
H_z=k\sqrt{D^2-E^2}, \qquad k=0,\pm 1, \pm 2,\ldots
\end{equation}
should be used instead of Eq.\ (\ref{ResCond}).
Although for $E\ll D$ the corrections to the resonance fields $H_z$ due to the transverse
anisotropy are small, the detuning of levels arising from the use of the approximate resonance
condition of Eq.\ (\ref{ResCond}) is much larger than the level splitting.
If the exact Eq.\ (\ref{ResCondExact}) was not known (which may well be the case for more
complicated spin models), the numerical finding of level splittings would require sweeping of
$H_z$ which is a time-costly procedure.

One of the results of this section is that $S\gg 1$ the slowly decaying perturbations of the
spin Hamiltonian due to dislocations, Eq.\ (\ref{EHtrGen}), generate tunnel splittings which
change by many orders of magnitude on the distance from the dislocation.
Consequently, the behavior of Mn$_{12}$ crystals at low temperatures must be similar to that
of disordered systems with widely distributed parameters.
The natural scale to discuss physical properties of such systems is logarithmic, which makes
the results practically insensitive to any prefactors of order unity, such as the functions
$g(\varphi)$ in Eq.\ (\ref{EHtrGen}) encapsulating differences between different types of
linear dislocations.
For this reason, we will use only one type of dislocations for illustrations, the edge
dislocations along the $y$-axis.
Before studying spin-tunneling rates due to dislocations in real Mn$_{12}$ crystals, one
should work out one fundamental characteristic of these crystals: the spatial distribution of
transverse anisotropies.
This will be done in the next section.

\section{Distribution of transverse anisotropies due to
dislocations} \label{sec_random}

In a crystal with dislocations, the deformation tensor at any given point is a sum of
contributions due to many different dislocations.
The superposition principle for deformations follows from the linearity of the equations of
the theory of elasticity \cite{lanlif7} and it holds everywhere outside dislocation cores,
i.e., for the distances from the dislocation axes $r \gtrsim 1$.
Statistical properties of deformations and thus of the spin tunneling rates in a Mn$_{12}$
crystal depend on the spatial distribution of dislocations which is poorly known.
In this section we present analytical solution of the problem of distribution of transverse
anisotropies, assuming that dislocations are distributed at random.
Note that alternatively, one can consider a regular array of dislocations with alternating
directions of the Burgers vector, to achieve a balance of contractions and dilations
throughout the crystal.
Such a dislocation array should be randomized to some extent to make it more realistic, as was
done in Ref.\ \onlinecite{garchu01prl}.

For the random-dislocation model, one should, in principle, consider different types of
dislocations and compute an average over all crystal sites.
For simplicity, we will take into account only one type of randomly distributed dislocations
with collinear axes: the edge dislocations along the $Y$-axis, see Sec.\ \ref{sec_couplings}E,
and consider one representative site in the middle of the crystal, ${\bf r}=0$.
The distribution function for the transverse anisotropy can be defined as
\begin{equation}\label{fEDef}
f_{\tilde E} = \left\langle \delta \left(\tilde E - \sum_{i=1}^N \tilde E({\bf r}_i) \right)
\right\rangle, \qquad \tilde E \equiv \frac E {2D},
\end{equation}
where $N \gg 1$ is the number of dislocations in the crystal and the averaging is carried out
over their positions ${\bf r}_i$ in the plane perpendicular to the dislocation axis within a
circular region of radius $R$.
One can define
\begin{equation}\label{RcDef}
c = \frac N {\pi R^2} = \frac 1 {\pi R_c^2},
\end{equation}
where $c$ is the concentration of dislocations and $R_c$ is the characteristic distance
between dislocations.

Let us at first analyze the large-$|\tilde E|$ asymptotes of $f_{\tilde E}$ due to the regions
with large deformations of both signs close to one of dislocations.
In that case one can neglect the influence of all other dislocations and consider the
one-dislocation model
\begin{equation}\label{fEoneDef}
f_{\tilde E} =  \frac 1 {\pi R_c^2} \int_0^{2\pi} d\varphi \int_0^{R_c} rdr \delta
\left(\tilde E - \frac{g(\varphi)} r \right).
\end{equation}
Integration yields
\begin{equation}\label{fEoneRes}
f_{\tilde E} = \frac{ \tilde E_c^2} { |\tilde E|^3 }, \qquad |\tilde E| \gtrsim \tilde E_c
\equiv \frac{ \sqrt{\langle g(\varphi)^2\rangle}}{R_c},
\end{equation}
where $E_c$ is the characteristic transverse anisotropy at the distance $R_c$.
This formula becomes invalid for $\tilde E \lesssim \tilde E_c$, where the lines of constant
$\tilde E$ in Eq.\ (\ref{fEoneDef}) cross the boundary of the region under consideration,
$r=R_c$.
In fact, for $\tilde E \lesssim \tilde E_c$ the very Eq.\ (\ref{fEoneDef}) becomes invalid and
one has to take into account other dislocations.
Eq.\ (\ref{fEoneRes}) suggests that one should introduce the distribution function for a
reduced transverse anisotropy $\alpha$
\begin{equation}\label{falphaDef}
f_\alpha \equiv \tilde E_c f_{\tilde E}, \qquad \alpha \equiv\tilde E/\tilde E_c,
\end{equation}
which has the asymptote
\begin{equation}\label{falphaAsymp}
f_\alpha = \frac 1 {|\alpha|^3}, \qquad \alpha \gtrsim 1.
\end{equation}

\begin{figure}[t]
\unitlength1cm
\begin{picture}(11,6.5)
\centerline{\psfig{file=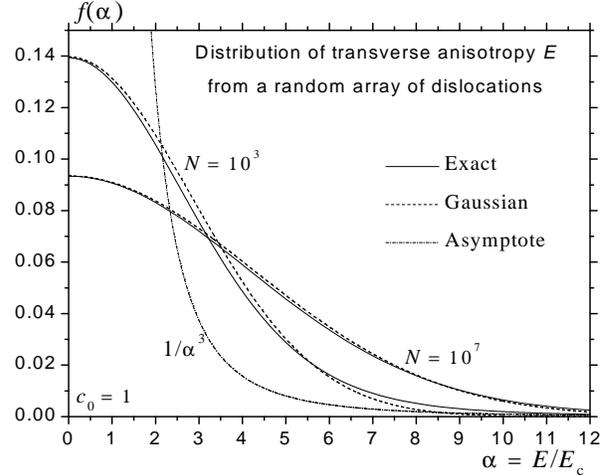,angle=-90,width=9cm}}
\end{picture}
\caption{ \label{fig_distribe} Distribution of the transverse anisotropy due to a random array
of edge dislocations along the $Y$-axis}
\end{figure}

In the general case, with the help of the identity $2\pi\delta(x) =
\int_{-\infty}^{\infty}d\omega e^{i\omega x}$, the averaging over the coordinates of different
dislocations in Eq.\ (\ref{fEDef}) can be factorized,
\begin{equation}\label{fEFact}
f_{\tilde E} = \int_{-\infty}^{\infty} \frac{d\omega}{2\pi} e^{i\omega \tilde E} f(\omega)^N,
\end{equation}
where
\begin{equation}\label{fomegaDef}
f(\omega) \equiv  \frac 1 {\pi R^2} \int_0^{2\pi} d\varphi \int_0^{R} rdr \exp\left(
-\frac{i\omega g(\varphi)} r \right).
\end{equation}
As we shall see, in Eqs.\ (\ref{fEFact}) and (\ref{fomegaDef}), $\omega \sim R_c \ll R$ for
$N\gg 1$, thus the argument of the exponential in Eq.\ (\ref{fomegaDef}) is small and
$f(\omega)$ is close to unity.
Then the exponential can be expanded and integrated, with a log accuracy, in the interval
$|\omega| \lesssim r < R$.
Given that $\langle g(\varphi)\rangle=0$, the result has the form
\begin{equation}\label{fomega}
f(\omega) \cong  1 - \frac{ \omega^2 \langle g(\varphi)^2\rangle }{R^2} \ln\frac{c_0
R}{|\omega| \sqrt{\langle g(\varphi)^2\rangle}},
\end{equation}
where $c_0$ is a constant of order unity.
Now with the use of Eqs.\ (\ref{RcDef}) and (\ref{fEoneRes}) one can write
\begin{equation}\label{fomegaN}
f(\omega)^N \cong  \exp\left[-(\omega \tilde E_c)^2 \ln\frac{c_0\sqrt{N}}{|\omega| \tilde
E_c}\right].
\end{equation}
At this point one may forget about the initial assumption on the circular form of the spatial
region.
The shape of the crystal only affects the value of the constant $c_0$ under the logarithm.
Eq.\ (\ref{fomegaN}) confirms the assumption $\omega \sim 1/\tilde E_c \sim R_c$ made above.
Now we are prepared to write down the final result which is convenient to formulate in terms
of the function $f_\alpha$ defined by Eq.\ (\ref{falphaDef})
\begin{equation}\label{falphaRes}
f_\alpha \cong \frac 1 \pi \int_0^\Lambda du \cos(\alpha u) \exp\left( -u^2 \ln \frac {c_0
\sqrt{N}}u \right).
\end{equation}
Here the cutoff $\Lambda$ satisfies $1 \ll \Lambda \ll \sqrt{N}$; one cannot integrate up to
$\infty$ since the form if the integrand is only valid for $u \ll \sqrt{N}$.
Clearly, for large enough crystals with $N\gg 1$ the result does not depend on $\Lambda$.
We remind that for the edge dislocations along the $Y$-axis, the distribution of transverse
anisotropies is an even function.
This distribution is shown for $\tilde E > 0$ in Fig.\ \ref{fig_distribe}.

Integrating Eq.\ (\ref{falphaRes}) by parts three times, one can recover the asymptote of
$f_\alpha$ at $|\alpha|\gg 1$ which is given by Eq.\ (\ref{falphaAsymp}).
This power-law asymptote is a consequence of the logarithmic singularity of the integrand in
Eq.\ (\ref{falphaRes}) at $u\to 0$ and it leads to the divergence of the second moment of
$f_\alpha$.
On the other hand, for large $N$ the distribution function may be well approximated by
Gaussian for not too large $\alpha$.
Indeed, for large $N$ the logarithm in Eq.\ (\ref{falphaRes}) is weakly dependent on $u$ and
can be replaced by a constant.
The best value of this constant corresponds to $u$ for which the argument of the exponential
equals one.
This requires solving a transcendental equation that can be done in a perturbative way.
With a good accuracy one can use
\begin{equation}\label{LogValue}
\ln \frac {c_0 \sqrt{N}}u\Rightarrow L = \ln\left[c_0\sqrt{N\ln(c_0\sqrt{N})}\right]
\end{equation}
which results in the aproximation
\begin{equation}\label{Gaussian}
f_\alpha \cong \frac 1 {2\sqrt{\pi L}} \exp\left( - \frac {\alpha^2}{4L} \right)
\end{equation}
which is also shown in Fig.\ \ref{fig_distribe}.

As the number $N$ of dislocations in the crystal inreases, the function $f_\alpha$ of Eq.\
(\ref{falphaRes}) becomes closer and closer to the Gaussian, whereas the power-law asymptote
given by Eq.\ (\ref{falphaAsymp}) becomes shifted to the region of very large $\alpha$ where
it is hardly visible.
This effect is due to the accumulation of small contributions from dislocations situated at
large distances from the observation point (of order of the linear dimension of the crystal).
Such small contributions from distant dislocations, which lead to the Gaussian distribution
$f_\alpha$, win over contributions from close dislocations responsible for Eq.\
(\ref{falphaAsymp}).
One can estimate the characteristic value $\alpha_1$ of the transverse anisotropy at which the
distribution function changes its behavior by equating the Gaussian approximation for
$f_\alpha$ given by Eq.\ (\ref{Gaussian}) to its asymptote of Eq.\ (\ref{falphaAsymp}).
This yields
\begin{equation}\label{alpha1}
\alpha_1 \cong \sqrt{4L\ln(4L/\sqrt{\pi})}.
\end{equation}
Transverse anisotropies $\tilde E$ [see Eqs.\ (\ref{fEDef}) and (\ref{falphaDef})]
corresponding to $\alpha
> \alpha_1$ are due to a single dislocation in the vicinity of the observation point, whereas
$\alpha < \alpha_1$ are collective contributions of many distant dislocations.
The characteristic distance $r_1$  from the dislocation core is defined by $\tilde E_1 \equiv
\alpha_1 \tilde E_c = \sqrt{\langle g(\varphi)^2\rangle}/r_1$ and reads
\begin{equation}\label{r1}
r_1 = \frac {R_c}{\alpha_1} = \frac 1 { \sqrt{4\pi cL\ln(4L/\sqrt{\pi})} }.
\end{equation}
One can see that in a macroscopic crystal the contribution of a particular dislocation
dominates within a distance from its core, $r_1$, that is small compared to the average
distance between the dislocations.
The fraction of cites in the lattice affected mostly by one close dislocation is
\begin{equation}\label{n1}
n_1 \cong 2\int_{\alpha_1}^\infty \frac{ d\alpha }{\alpha^3} = \frac 1 {\alpha_1^2} = \frac 1
{4L\ln(4L/\sqrt{\pi})}.
\end{equation}
This fraction is small if the number of dislocations $N$ in the crystal is large.
Molecules belonging to this group may be interpreted as the minor species.
One should note that the value of $n_1$ above is, in fact, only the upper bound on $n_1$.
The asymptote $f_\alpha \cong 1/\alpha^3$ is applicable for $\alpha \gg \alpha_1$ (say, for
$\alpha \gtrsim 3\alpha_1$) rather than for $\alpha \gtrsim \alpha_1$.
If one replaces $ \alpha_1$ by $3\alpha_1$ in Eq.\ (\ref{n1}), the value of $n_1$ will
decrease by one order of magnitude.

The Gaussian approximation for the function $f_{\tilde E}$ with the help of Eq.\
(\ref{falphaDef}) can be written in the form
\begin{equation}\label{GaussianE}
f_{\tilde E} \cong \frac 1 {2\tilde E_{\tilde c} \sqrt{\pi}} \exp\left( - \frac {\tilde
E^2}{(2\tilde E_{\tilde c})^2} \right),
\end{equation}
where
\begin{equation}\label{EctilDef}
\tilde E_{\tilde c} \equiv \tilde E_c \sqrt{L} = \sqrt{\pi \langle g(\varphi)^2\rangle \tilde
c}, \qquad \tilde c \equiv cL.
\end{equation}
That is, the accumulation of contributions from distant dislocations leads to the effective
logarithmic renormalization of the concentration of dislocations $c$ with $L$ defined by Eq.\
(\ref{LogValue}).
For edge dislocations along the $Y$-axis, the quantity $\sqrt{ \langle g(\varphi)^2\rangle}$
according to Eqs.\ (\ref{EHtrGen}), (\ref{EHxDef}), and (\ref{xxyyDefEdgeY}) is given by
\begin{equation}\label{gAvrEdgeY}
\sqrt{ \langle g(\varphi)^2\rangle} = \frac{g_1 \sqrt{5-12\sigma+8\sigma^2}} {8\pi(1-\sigma)},
\end{equation}
where $\sigma$ is the Poisson elastic coefficient.
For for $g_1=1$ and $\sigma=0.25$ one has $\sqrt{ \langle g(\varphi)^2\rangle}\approx 0.084$.

The experimentally studied Mn$_{12}$ crystals are rather large, about $0.5 \times 0.5$ mm$^2$,
which corresponds to the cross-section of about $10^{11}$ lattice cells.
Even for the concentration of dislocations as small as $c=10^{-4}$ per cell, the number of
dislocation in the crystal is about $N\approx 10^7$.
For $c_0=1$ this gives $L=9.1$, i.e., the effective concentration of dislocations increases by
an order of magnitude, $\tilde c = 0.91\times 10^{-3}$.
The corresponding value of $\tilde E_{\tilde c}$ that follows from Eqs.\ (\ref{EctilDef}) and
(\ref{gAvrEdgeY}) is $\tilde E_{\tilde c}= 0.449\times 10^{-2}$.
For $c=10^{-3}$ one obtains $L=10.3$, thus $\tilde E_{\tilde c}= 1.51\times 10^{-2}$.
If one takes $L\approx 10$, then Eq.\ (\ref{n1}) yields $n_1 \approx 1/125$.
That is, less than 1\% of all  Mn$_{12}$ molecules in the crystal belong to the fast relaxing
group in the vicinity of a single dislocation.
In fact, the value of $n_1$ should be much smaller, see comment after Eq.\ (\ref{n1}).
The renormalization of the concentration of dislocations and the Gaussian distribution of
transverse anisotropies for large crystals are clearly seen in Fig.\ \ref{fig_distribe}: The
distribution becomes broader in the $\alpha$-scale due to the increase $L$ with $N$, Eq.\
(\ref{LogValue}).

\begin{figure}[t]
\unitlength1cm
\begin{picture}(11,6.5)
\centerline{\psfig{file=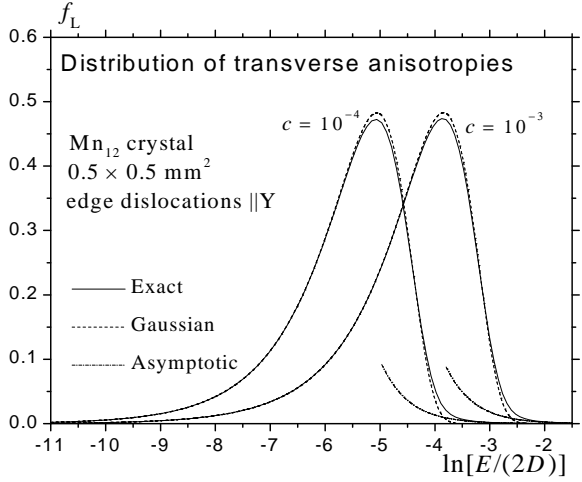,angle=-90,width=9cm}}
\end{picture}
\caption{ \label{fig_distrelog} Distribution of the logarithm of transverse anisotropy due to
random array of edge dislocations along the $Y$-axis}
\end{figure}

Having determined the distribution of transverse anisotropies created by dislocations in a
Mn$_{12}$ crystal, we can calculate the distribution of tunnel splittings for different level
pairs and different resonances.
This will be done in the next section, where we will use the distribution of the logarithm of
the transverse anisotropy
\begin{equation}\label{fLDef}
f_L(x) = 2 e^x f_{\tilde E}(e^x), \qquad x\equiv \ln \tilde E .
\end{equation}
If the Gaussian approximation for $f_{\tilde E}$, Eq.\ (\ref{GaussianE}), is adopted, then
$f_L$ has the form
\begin{equation}\label{GaussianEL}
f_L(x) \cong \frac 1 {\tilde E_{\tilde c} \sqrt{\pi}} \exp\left(x - \frac {e^{2x}}{(2\tilde
E_{\tilde c})^2} \right).
\end{equation}
This function is plotted in Fig.\ \ref{fig_distrelog} for $c=10^{-3}$ ($\tilde E_{\tilde
c}\approx 1.51\times 10^{-2}$) and $c=10^{-4}$ ($\tilde E_{\tilde c}\approx 0.449\times
10^{-2}$).

\section{Distribution of tunnel splittings}
\label{sec_distrsplittings}

\begin{figure}[t]
\unitlength1cm
\begin{picture}(11,6.5)
\centerline{\psfig{file=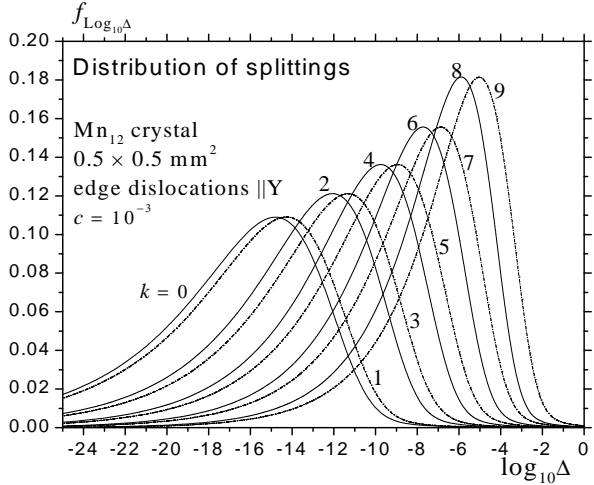,angle=-90,width=9cm}}
\end{picture}
\caption{ \label{fig_distrslog} Distribution of the logarithm of the ground-state ($n=0$)
tunnel splittings due to a random array of edge dislocations along the $Y$-axis}
\end{figure}

Distribution of transverse anisotropies due to dislocations, which was obtained in the
previous section, determines the distribution of the tunnel splittings which for the even-$k$
resonances are given by Eq.\ (\ref{splitE}).
It is convenient to rewrite Eq.\ (\ref{splitE}) in terms of the resonance number $k$ and the
level number $n=0, 1,\ldots$ in the metastable well of negative spin projections
\begin{equation}\label{nkDef}
m=n-S, \qquad m'=S-n-k.
\end{equation}
Thus Eq.\ (\ref{splitE}) becomes
\begin{equation}\label{SplEvenRewr}
\Delta_{nk} = g_{nk} \tilde E ^{\xi_{nk}} \qquad (k \;\;{\rm even}),
\end{equation}
where $\tilde E\equiv E/(2D)$, $\xi_{nk}\equiv S-n-k/2$, and
\begin{equation}\label{gnkDef}
g_{nk} \equiv \frac{2D}{[(2S-2n-k-2)!!]^2}
 \sqrt{\frac{ (2S-n-k)! (2S-n)! }{ (n+k)! n! } }.
\end{equation}
With the help of Eq.\ (\ref{SplEvenRewr}) one can write down the distribution of the square of
tunnel splittings for any resonance in terms of the previously defined $f_{\tilde E}$
\begin{equation}\label{fDelta2}
f_{\Delta_{nk}^2}(\Delta^2) = \frac 1 {\xi_{nk}g_{nk}^2} \left[\frac
{\Delta^2}{g_{nk}^2}\right]^{\frac 1{2\xi_{nk}}-1} \!\!f_{\tilde E}\left[\left(\frac
{\Delta^2}{g_{nk}^2}\right)^\frac 1{2\xi_{nk}}\right].
\end{equation}
Because $\xi_{nk}$ is a large number, especially for resonances with small $n$ and $k$, tunnel
splittings in a Mn$_{12}$ crystal spead over many orders of magnitude, from rather large
values near dislocation cores to very small values far from dislocations.
It is thus more convenient to consider the distribution of the decimal logarithm of tunnel
splittings, $y\equiv {\rm log}_{10}\Delta$.
With the help of Eq.\ (\ref{SplEvenRewr}) it can be written as
\begin{equation}\label{SplDistrEven}
f_{{\rm log}_{10}\Delta} (y) = \frac {\ln10} {\xi_{nk}} f_L \left( \frac {y\ln10 -\ln g_{nk}}
{\xi_{nk}} \right).
\end{equation}
Here $f_L(x)$ is the distribution function of $x=\ln \tilde E$ defined by Eq.\ (\ref{fLDef}).
Distribution of the ground-state ($n=0$) tunnel splittings for different resonance numbers $k$
is shown in Fig.\ \ref{fig_distrslog} for a Mn$_{12}$ crystal of size $0.5 \times 0.5$ mm$^2$
with concentration of dislocations $c=10^{-3}$.
One can see that the distribution is shifted to the left and becomes broader for smaller $k$.

For odd tunneling resonances, the splitting, according to Eq.\ (\ref{splitEHx}), depends on
both the transverse anisotropy and the transverse field which are generated by dislocations.
Since the functions $g(\varphi)$ and $g_H(\varphi)$ in Eq.\ (\ref{EHtrGen}) are different, one
cannot, strictly speaking, express the distribution of splittings via that of the transverse
anisotropy alone.
Fortunately, transverse field is only a prefactor in front of a small term in Eq.\
(\ref{splitEHx}).
This allows one to express $H_\perp$ via $E$ with the help of Eq.\ (\ref{EHtrGen}) and use the
resonance condition $H_z=kD$ to obtain
\begin{equation}\label{HxviaE}
\frac{H_x}{2D} = \frac{k\tilde E} 2 \frac{g_H(\varphi)}{g(\varphi)} \Rightarrow \frac{k\tilde
E} 2 c_{\rm odd},
\end{equation}
with the ratio of angular functions approximated by a constant $c_{\rm odd}\sim 1$.
In this approximation, the level splittings at odd $k$ are functions of the transverse
anisotropy alone, so that one can rewrite Eq.\ (\ref{splitEHx}) as
\begin{equation}\label{SplOddRewr}
\Delta_{nk} = g'_{nk} \tilde E ^{\xi'_{nk}} \qquad (k \;\;{\rm odd}),
\end{equation}
where
\begin{equation}\label{gprDef}
g'_{nk} \equiv \frac{ck}2 g_{nk}, \qquad \xi'_{nk} = \xi_{nk} + \frac 12 = S-n-\frac{k-1}2
\end{equation}
[cf.\  Eqs.\ (\ref{SplEvenRewr}) and (\ref{gnkDef})].
For the distribution of the decimal logarithm of splittings at odd resonances one obtains the
formula analogous to Eq.\ (\ref{SplDistrEven}) where the constant $c$ enters under the
logarithm.
Note that the power $\xi'_{nk}$ for an odd resonance is the same as the power $\xi_{nk}$ for
the preceding even resonance, thus the splitting distribution functions for these two
resonances differ, according to Eq.\ (\ref{SplDistrEven}), only by a shift.
This is clearly seen on Fig.\ \ref{fig_distrslog} where we have plotted the odd-resonance
curves for $c_{\rm odd}=1$.

If the distribution of the level splittings for different $k$ is simulated, as was done in
Ref.\ \onlinecite{garchu01prl}, or is measured, the theory developed above could be tested by
re-plotting the data in terms of the scaling variable\cite{garchu01prl}
\begin{equation}\label{Scaling}
x = \frac 1{\xi_{nk}} \ln\frac{\Delta_{nk}}{g_{nk}}.
\end{equation}
Since $x=\ln\tilde E$, it does not depend on the numbers $n$ and $k$, thus all the curves
should scale.
For odd resonances, one should use $\xi'_{nk}$ and $g'_{nk}$ of Eq.\ (\ref{gprDef}), where $c$
can be considered as an adjustable parameter.

\section{Incoherent Landau-Zener processes}
\label{sec_landauzener}

A very convenient method to experimentally study spin tunneling consists of sweeping the
longitudinal field, thus making a pair of levels on different sides of the barrier,
$\varepsilon_n$ and $\varepsilon_{n'}$, to go through a resonance. \cite{wersesgat99,werses99}
For a purely quantum-mechanical system, the probability of the system to stay in the same well
(i.e., to go from one {\em exact} energy branch to the other) is given by
\cite{lan32,zen32,dobzve97}
\begin{equation}\label{LanZenQ}
P = \exp\left( - \frac{ \pi \Delta_n^2 } { 2v_n } \right ),
\end{equation}
where $\Delta_n$ is the level splitting and $v_n \equiv |d(\varepsilon_{n'}
-\varepsilon_n)/dt|$ is the speed of the level detuning {\em unperturbed} by the tunneling
interaction, i.e., the energy sweep rate.
Landau\cite{lan32} has obtained this formula in the special case of a large argument of the
exponential (low sweep rates), whereas Zener\cite{zen32} obtained it for all sweep rates,
however with a wrong additional factor $2\pi$ in the exponential.
Recent Ref.\ \onlinecite{dobzve97} reproduces Zener's results without the wrong factor $2\pi$
[Eq.\ (11) of Ref.\ \onlinecite{dobzve97} is equivalent to Eq.\ (\ref{LanZenQ})].
We will call this process the {\em coherent} Landau-Zener process.

For systems interacting with the environment, the situation changes if the linewidth
$\Gamma_n$ of the level is greater than the level splitting $\Delta_n$.
This, in fact, happens in most cases, since $\Delta_n$ for spin systems is a high power of a
small perturbation, while $\Gamma_n$ is due to transitions between the levels in the same well
which appear in much lower order of the perturbation theory on spin-phonon (or other)
interactions.
The only exception is the tunneling resonance between the ground-state levels in the two wells
at low temperatures because the phonon processes contributing to their linewidths die out
exponentially at $T\to 0$.
If at least one of the resonant levels is not the lowest level in the well, it has a
considerable linewidth down to $T=0$ due to the transitions onto lower levels in the same
well, accompanied by the emission of phonons.
In this case $\Gamma_n \gg \Delta_n$ and tunneling becomes mediated by the environment, i.e.,
incoherent.

Incoherent spin tunneling should be described in terms of the density matrix equation rather
than in terms of the Schr\"odinger equation.\cite{garchu97}
Since at low temperatures the escape process (via tunneling or via thermal activation) is much
slower than the equilibration within the wells, one can eliminate the non-diagonal elements of
the density-matrix equation and obtain the system of equations for the level populations that
describes tunneling, Eq.\ (4.12) of Ref.\ \onlinecite{garchu97}.
In sweeping-field experiments on Mn$_{12}$ in the kelvin or subkelvin range, the backflow of
particles from the stable to the metastable well is exponentially small and can be safely
neglected for all resonances with $k\neq 0$, since the corresponding activation energy is
about $2SD\simeq 13$K.
Thus there is only one-way escape from the metastable well described by the
equation\cite{garchu97}
\begin{equation}\label{KinEq}
\dot N_0 = - \sum_{n=0}^{n_{\rm top}} N_n \frac{\Delta_{nk}^2}2 \frac{ \Gamma_{nn'} }{
(\varepsilon_{n'} -\varepsilon_n)^2 + \Gamma_{nn'}^2 },
\end{equation}
where $\Gamma_{nn'} \equiv \Gamma_{n}+\Gamma_{n'}$, the numbering of levels begins from the
metastable ground state, $n=m+S$, and $n_{\rm top}$ corresponds to the top of the barrier.
The tunnel splitting $\Delta_{nk}$ depends on both the level number $n$ in the metastable well
and the number of the resonance $k$ of Eq.\ (\ref{ResCond}).
They are given by Eqs.\ (\ref{splitE}) and (\ref{splitEHx}) with $m=n-S$ and $m'=S-n-k$.
In the kelvin and subkelvin temperature range almost all Mn$_{12}$ molecules in the metastable
well are in the ground state, whereas the populations of the excited states entering Eq.\
(\ref{KinEq}) are exponentially small and are given by the equilibrium formulas
\begin{equation}\label{NnEquil}
N_n = N_0 e^{-\Delta \varepsilon_{nk} /T}, \qquad
 \Delta \varepsilon_{nk} = \varepsilon_{n} - \varepsilon_{0},
\end{equation}
where the level energies are taken at the resonant values of the field, $H_z=kD$ ($k\geq 0$)
\begin{equation}\label{epsnk}
 \varepsilon_{n} = Dn(2S-n-k) -DS(S-k).
\end{equation}
Now Eq.\ (\ref{KinEq}) can be integrated for the field sweeping across the resonance.
For the constant energy-sweep rate
\begin{equation}\label{EnergySweep}
\varepsilon_{n'} -\varepsilon_n = v_{nk}t + {\rm const},
\end{equation}
where $v_{nk}=(2s-2n-k)dH_z/dt$, one obtains
\begin{equation}\label{NBeforeAfter}
N_0^{(k,\rm after)} \!= N_0^{(k,\rm before)} \!\exp\!\left[ - \sum_{n=0}^{n_{\rm max}} \frac {
\pi \Delta_{nk}^2  e^{-\Delta \varepsilon_{nk}/T} } {2v_{nk} } \right],
\end{equation}
where $N_0^{(k,\rm before)}$ and $N_0^{(k,\rm after)}$ are the numbers of Mn$_{12}$ molecules
in the metastable well before and after crossing the $k$-resonance.

Eq.\ (\ref{NBeforeAfter}) describes a superposition of incoherent Landau-Zener tunneling
processes which come both directly from the metastable ground state and via excited states.
At $T=0$ the result coincides with that of Eq.\ (\ref{LanZenQ}), although the physics is
different.
Note that for the incoherent Landau-Zener tunneling between the two lowest-energy states one
has to take into account the flow of particles in both directions, which results in the
disappearance of the factor 2 in the denominator of Eq.\ (\ref{NBeforeAfter}).
\cite{leulos00zen}
In this situation, however, one should carefully check the applicability condition of the
method, $\Gamma_n \gg \Delta_n$, since $\Gamma_n$ may be very small.
Incidentally, for Mn$_{12}$ acetate the tunneling between the ground states is so weak that
the $k=0$ resonance cannot be observed experimentally, unless a strong transverse field is
applied.

The field sweeping technique is a very convenient experimental tool since Eq.\
(\ref{NBeforeAfter}) does not depend on the damping parameters which are difficult to
determine with sufficient accuracy.
The combination $\Delta_{nk}^2 e^{-\Delta \varepsilon_{nk} /T}$ is familiar from Refs.\
\onlinecite{garchu97,garchu99,garchu00} where it was used to investigate the transition on
temperature between thermally activated and tunneling regimes of the escape from the
metastable well.
To be more precise, let us rewrite Eq.\ (\ref{NBeforeAfter}) with the help of
\begin{eqnarray}\label{SweepRate}
&& v_{nk} = \delta \varepsilon_{nk} \tilde v, \qquad \tilde v \equiv \frac 1 D \frac
{dH_z}{dt}
\nonumber\\
&&
 \delta \varepsilon_{nk} \equiv \varepsilon_n - \varepsilon_{n-1} = D(2S-2n-k)
\end{eqnarray}
where $\delta \varepsilon_{nk}$ is the level spacing in the well.
One obtains
\begin{equation}\label{NBeforeAfterRewr}
N_0^{(k,\rm after)} = N_0^{(k,\rm before)} \exp\left[ - \sum_{n=0}^{n_{\rm max}} \frac { \pi
f_{nk}(T) } {2\tilde v } \right],
\end{equation}
where
\begin{equation}\label{fnkDef}
f_{nk}(T) \equiv \frac{\Delta_{nk}^2}{\delta \varepsilon_{nk} } e^{-\Delta
\varepsilon_{nk}/T}.
\end{equation}
In fact, this form of $f_{nk}$ is valid when the tunnel splitting is small, $\Delta_{nk} \ll
\delta \varepsilon_{nk}$.
This is true for energy levels below the top of the barrier.
Near the top of the barrier, where $\Delta_{nk} \sim \delta \varepsilon_{nk}$, a more accurate
formula is \cite{garchu99}
\begin{equation}\label{fnkDefImproved}
f_{nk}(T) \equiv \frac{\delta \varepsilon_{nk} \Delta_{nk}^2}{(\delta \varepsilon_{nk})^2 +
(\pi \Delta_{nk})^2 } e^{-\Delta \varepsilon_{nk}/T}.
\end{equation}
It ensures that $f_{nk} \sim \delta \varepsilon_{nk}  e^{-\Delta \varepsilon_{nk}}$ near the
top of the barrier for any approximation used for $\Delta_{nk}$.

Since $f(n)$ is a product of the two functions, one of which rapidly increases and the other
rapidly decreases as $n$ goes up, there is a narrow group of levels around $n=n^*(T)$ that
maximizes $f_{nk}(T)$ and makes the dominant contribution to the Landau-Zener transition.
Above the quantum-classical transition temperature $T_0$, one has $n^*(T)=n_{\rm top}$, which
corresponds to the activated regime with the spin escaping over the top of the barrier.
As $T$ is lowered below $T_0$ the tunneling level goes down, which corresponds to the
thermally assisted tunneling or, at even lower temperatures, to the ground-state tunneling.
If $n^*(T)$ is continuous, one can speak about the second-order quantum-classical transition.
If $n^*(T)$ jumps over several levels at some temperature, the transition is first
order.\onlinecite{garchu97,garchu99,garchu00}
Recently an experimental evidence of the first-order transition in Mn$_{12}$ has been obtained
in Refs.\ \onlinecite{bokkenwal00,kenetal00}.
(The biaxial model with the magnetic field along the hard axis shows two first-order
transitions in some range of parameters.\cite{marchu00,garchu00})

Experimental study of the temperature dependence of the tunneling level $n^*(T)$ in Mn$_{12}$
is possible due to the small quartic-anisotropy term $-AS_z^4$ in the spin Hamiltonian, which
makes the resonance fields $H_z$ dependent of the quantum number $n$.
Tunneling resonance between higher pairs of levels takes place at slightly lower value of
$H_z$ than the resonance between lower pairs of levels, that is, the $k$-resonance has a fine
structure.
Measuring temperature dependence of the resonance field $H_z$ for a given $k$ gives the
information on the level $n$ that dominates tunneling at a given
temperature.\cite{bokkenwal00,kenetal00}
Eq.\ (\ref{NBeforeAfter}) remains correct if the resonances for different $n$ take place at
different values of the field $H_z$, provided that the field is being swept through all these
resonances.
If different $n$-resonances are well separated from each other, which is the case for
Mn$_{12}$, then for each $n$-resonance one obtains
\begin{equation}\label{NnBeforeAfter}
N_0^{(n,k,\rm after)} = N_0^{(n,k,\rm before)} \exp\left[ -
 \frac { \pi \Delta_{nk}^2 e^{-\Delta
\varepsilon_{nk} /T} } {2v_{nk} } \right]
\end{equation}
[cf.\ Eq.\ (\ref{NBeforeAfter})].

\section{Landau-Zener relaxation in systems with distributed parameters}
\label{sec_sweeping}

As follows from the above, in Mn$_{12}$ crystals with dislocations the tunnel splittings
$\Delta_{nk}$ differ from one Mn$_{12}$ molecule to another.
Thus the fraction of Mn$_{12}$ molecules that stay in the metastable well after crossing the
$n$-resonance,
\begin{equation}\label{RnkDef}
R_{nk} \equiv \frac{N_0^{(n,k,\rm after)}}{N_0^{(n,k,\rm before)}}
\end{equation}
is an average that must be computed with the help of the distribution function of Eq.\
(\ref{fDelta2})
\begin{equation}\label{Rnk}
R_{nk}(\tilde v) = \int_0^\infty d\Delta^2 f_{\Delta_{nk}^2}(\Delta^2) \exp\left[-\frac{ \pi
f_{nk}(\Delta,T) }{2\tilde v} \right],
\end{equation}
where $f_{nk}$ is given by Eq.\ (\ref{fnkDefImproved}).
As was argued after Eq.\ (\ref{fDelta2}), the distribution of splittings is so broad that the
natural scale to represent it is logarithmic.
On the logarithmic scale, one can replace the exponential by a step function, $e^x \Rightarrow
\theta(1-x)$.
This gives
\begin{equation}\label{RnkSimpl}
R_{nk}(\tilde v) \cong \int_0^{\Delta_{nk,v}^2} d\Delta^2 f_{\Delta_{nk}^2}(\Delta^2),
\end{equation}
where $\Delta_{nk,v}^2$ is the solution of the equation
\begin{equation}\label{DelnkDef}
\pi f_{\Delta_{nk}^2}(\Delta^2)/(2\tilde v) = 1
\end{equation}
for $\Delta^2$.
With the help of Eq.\ (\ref{fnkDefImproved}) we get
\begin{equation}\label{Delnkv}
\Delta_{nk,v}^2 = \left(\frac {\delta\varepsilon_{nk}} \pi \right)^2 \frac {\tilde v}{\tilde
v_{nk}^{\rm max} -\tilde v}, \qquad \tilde v < \tilde v_{nk}^{\rm max},
\end{equation}
and $\Delta_{nk,v}^2=\infty$ for $\tilde v \geq \tilde v_{\rm max}$, where
\begin{equation}\label{vmaxdef}
\tilde v_{nk}^{\rm max} \equiv \frac {\delta\varepsilon_{nk}}{2\pi}e^{-\Delta
\varepsilon_{nk}/T}.
\end{equation}
\begin{figure}[t]
\unitlength1cm
\begin{picture}(11,6.5)
\centerline{\psfig{file=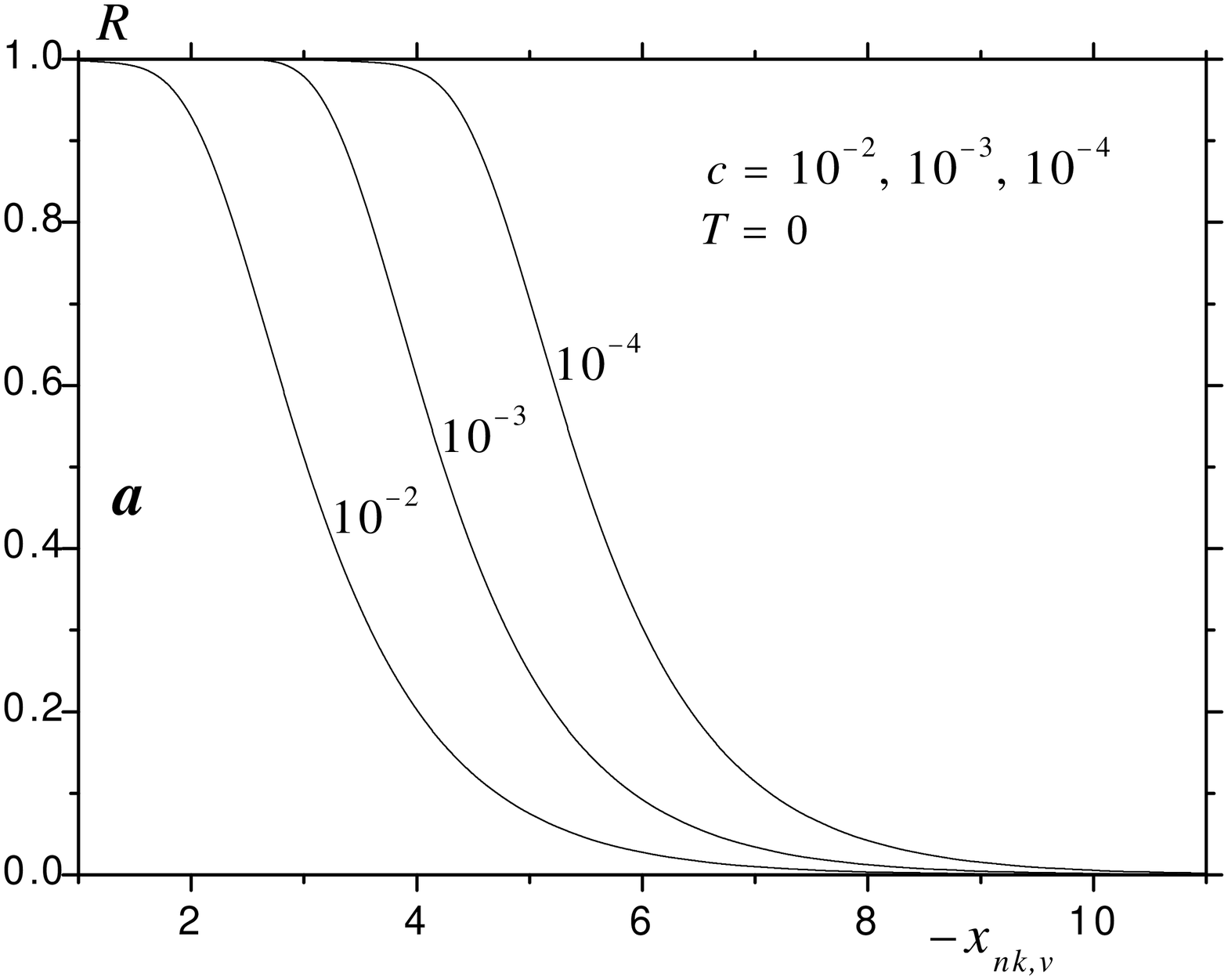,angle=-90,width=9cm}}
\end{picture}
\begin{picture}(11,5.8)
\centerline{\psfig{file=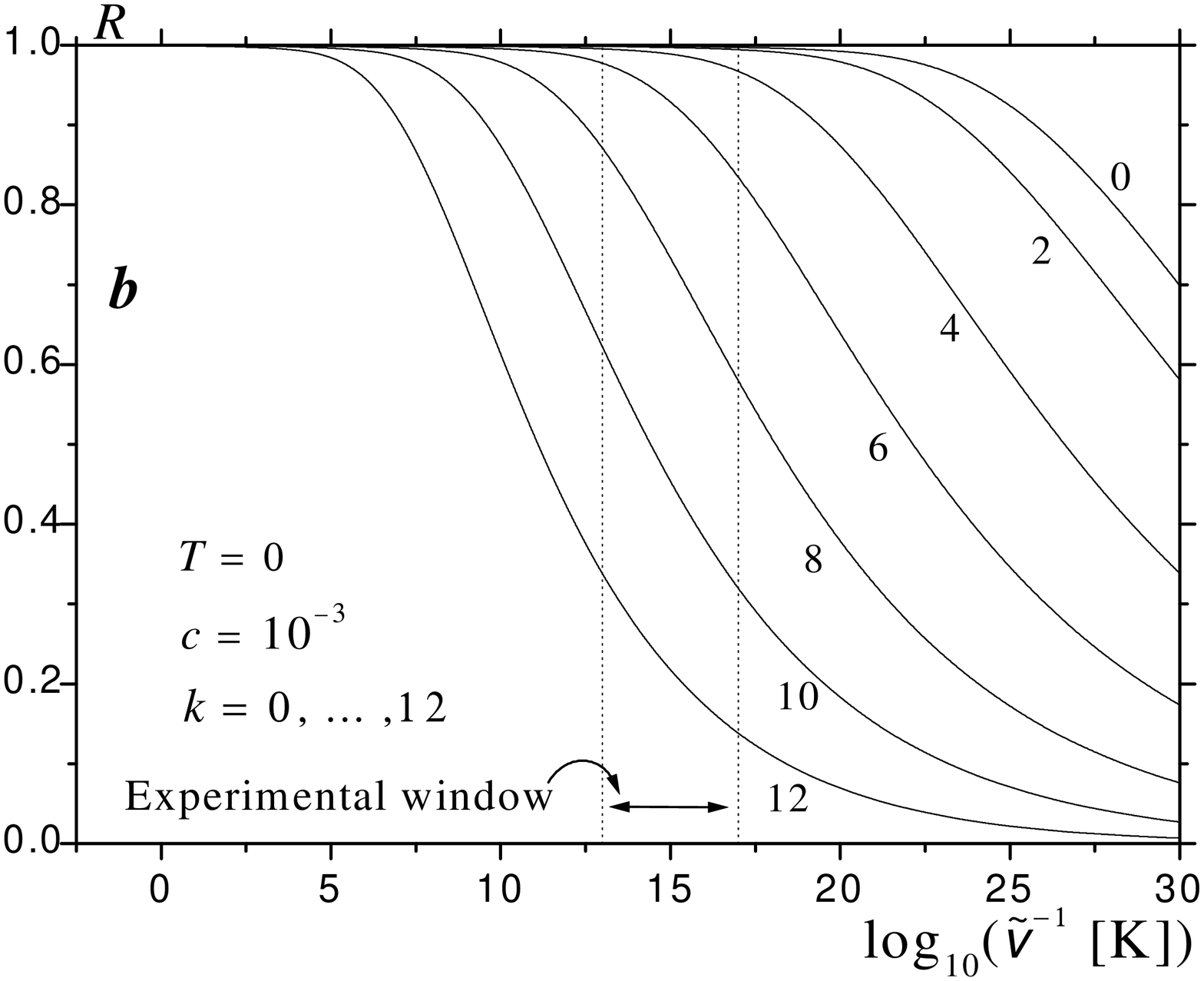,angle=-90,width=9cm}}
\end{picture}
\caption{ \label{fig_rel} $a$ -- Probability of remaining in the metastable well $R$ as a
function of the scaling argument $-x_{nk,v}$ [see Eq.\ (\protect\ref{xnkvDef})] at $T=0$; $b$
-- relaxation curves $R$ as functions of the sweep rate for different resonances $k$. }
\end{figure}

The physical meaning of Eq.\ (\ref{Rnk}) is the following.
Since the tunnel splitting $\Delta$ is a high power of the transverse anisotropy $E$, there is
very few Mn$_{12}$ molecules with $\Delta_{nk}\sim \Delta_{nk,v}$, whereas most of the
molecules have $\Delta_{nk}\ll \Delta_{nk,v}$ or $\Delta_{nk}\gg \Delta_{nk,v}$.
Eq.\ (\ref{Rnk}) simply ignores the small group of Mn$_{12}$ molecules with $\Delta_{nk}\sim
\Delta_{nk,v}$.
The quantity $\tilde v_{nk}^{\rm max}$ is the maximal sweep rate $\tilde v$ for which the
Landau-Zener transition for the given $n$, $k$, and $T$ is possible.
For $\tilde v \geq \tilde v_{nk}^{\rm max}$ one cannot find Mn$_{12}$ molecules that relax
fast enough to satisfy Eq.\ (\ref{Rnk}), since the transition probability has an upper bound
on $\Delta$ which is described by Eq.\ (\ref{fnkDefImproved}).
Eq.\ (\ref{RnkSimpl}) can be rewritten in terms of the distribution function for the logarithm
of the transverse anisotropy, Eq.\ (\ref{fLDef}).
It then becomes
\begin{equation}\label{RnkScaled}
R_{nk} \cong \int_{-\infty}^{x_{nk,v}} dx f_L(x),
\end{equation}
where $x_{nk,v}=\ln\tilde E_{nk,v}$ and $\tilde E_{nk,v}$ is the transverse anisotropy needed
to create the value of splitting $\Delta_{nk,v}$.
With the help of Eqs.\ (\ref{Delnkv}) and (\ref{SplEvenRewr}) we get
\begin{equation}\label{xnkvDef}
x_{nk,v} = \frac 1 {\xi_{nk}} \ln \left[\frac {\delta\varepsilon_{nk} }{\pi g_{nk}}
\sqrt{\frac{\tilde v}{\tilde v_{\rm max} -\tilde v } } \right]
\end{equation}
One can see from Eq.\ (\ref{RnkScaled}) that the relaxation curves $R_{nk}$ for the
$n,k$-resonance, when represented in terms of $x_{nk,v}$, depend only on the distribution of
transverse anisotropies in the crystal and scale onto the same curve.
If the Gaussian approximation for the anisotropy-distribution function is used, see Eqs.\
(\ref{GaussianE}) and (\ref{GaussianEL}), $R_{nk}(x_{nk,v})$ can be expressed via the error
function.
The relaxation curves at $T=0$ are shown in Fig.\ \ref{fig_rel} both in the scaling form, as
functions of $-x_v$, and in the natural form, as a function of ${\rm log}_{10}(1/\tilde v)$
for even $k$.
For odd $k$ the relaxation curves (not shown) differ by a shift from the ajacent even-$k$
curves, see Fig.\ \ref{fig_distrslog}.
The curves in Fig.\ \ref{fig_rel}b have been plotted with the help of the simplified Eq.\
(\ref{RnkSimpl}).
The deviation from the exact Eq.\ (\ref{Rnk}) is rather small and it can be further reduced if
one uses the improved formula
\begin{equation}\label{RnkScaledCorr}
R_{nk} \cong \int_{-\infty}^{x_{nk,v}} dx f_L(x) -\frac C {2\xi_{nk,v}} f_L(x_{nk,v}),
\end{equation}
where $C=0.577216$ is the Euler constant.
Note that the $1/\xi_{nk}$ correction above has a non-scaling form.
This correction is equivalent to a small shift of the relaxation curves in Fig.\ \ref{fig_rel}
to the left:
\begin{equation}\label{xnkvtil}
x_{nk} \Rightarrow \tilde x_{nk} = x_{nk} - C/(2\xi_{nk}).
\end{equation}
As we have seen in Sec.\ \ref{sec_distrsplittings}, distributions of tunnel splittings for any
$n$ and $k$ are functions of the distribution of transverse anisotropies, which is the basic
characteristic of the Mn$_{12}$ crystal.
Field-sweeping measurements in the subkelvin temperature range, where transitions via excited
levels, $n>0$, are negligible, provide the means of extracting that distribution.
For odd $k$-resonances, one should use Eq.\ (\ref{SplOddRewr}) and fit the constant $c_{\rm
odd}$.

\begin{figure}[t]
\unitlength1cm
\begin{picture}(11,11.1)
\centerline{\psfig{file=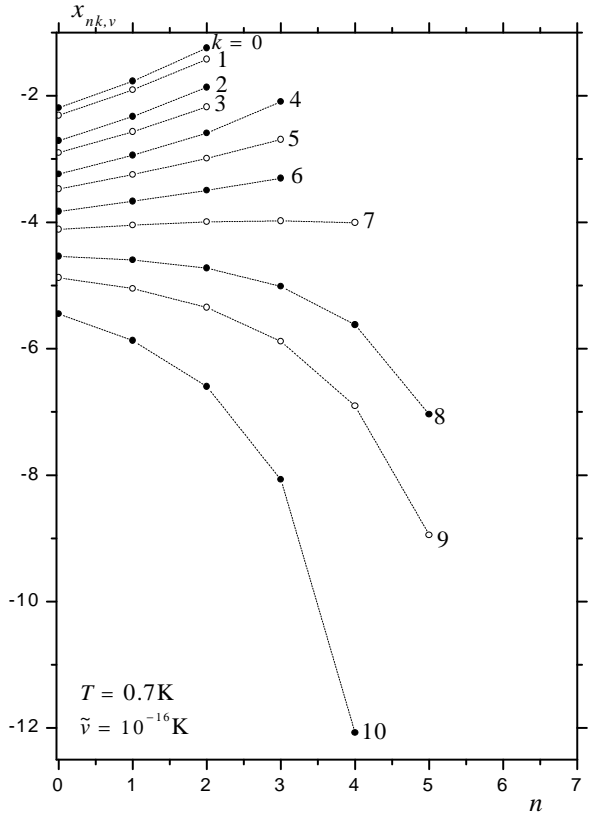,angle=0,width=9cm}}
\end{picture}
\caption{ \label{fig_xv07} Dependence of $x_{nk,v}=\ln \tilde E_{kn,v}$ [see Eq.\
(\protect\ref{xnkvDef})] on the excited level $n$ at different $k$ for $T=0.7$K and $\tilde v
= 10^{-16}$. }
\end{figure}
\begin{figure}[t]
\unitlength1cm
\begin{picture}(11,11.1)
\centerline{\psfig{file=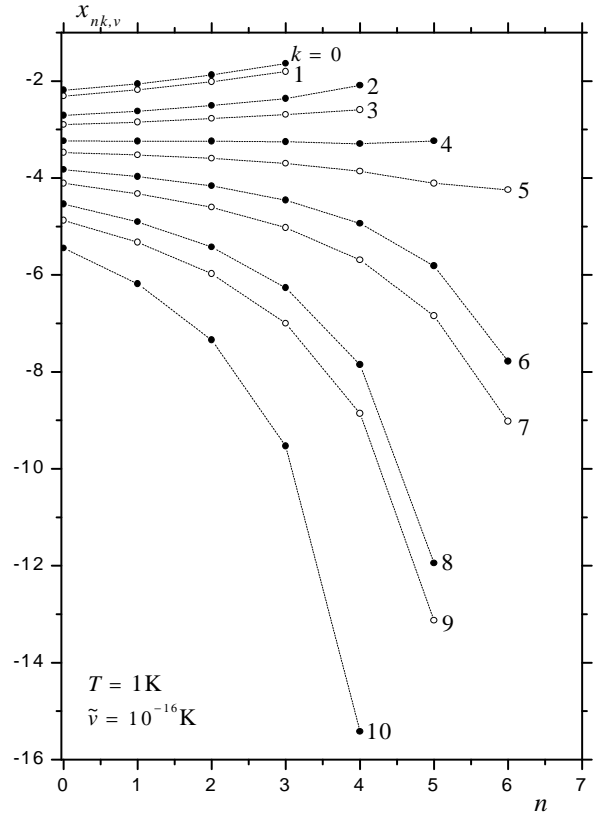,angle=0,width=9cm}}
\end{picture}
\caption{ \label{fig_xv1} The same  for $T=1$K. }
\end{figure}

At higher temperatures, tunneling transitions via excited states $n>0$ become activated.
The resonances with higher $n$ occur at lower values of the longitudinal field $H_z$ because
of the small term $-AS_z^4$ in the spin Hamiltonian.
In an ideal Mn$_{12}$ crystal, field-sweeping experiments could provide the information on the
level $n^*(T)$ which maximizes the function $f_{nk}$ and thus dominates the escape process at
temperature $T$.
Here the field sweep should not be too slow, otherwise crossing the resonances with higher $n$
will cause depletion of the metastable well and resonances with lower $n$ will not be seen.
In real Mn$_{12}$ crystals with dislocations, the situation changes drastically because of the
broad distribution of tunneling rates.
The depletion becomes unavoidable because for any $\tilde v < \tilde v_{nk}^{\rm max}$ most of
Mn$_{12}$ molecules in the crystal belong to one of two groups.
Molecules of the first group cross the barrier with a probability close to one, while
molecules of the second group stay in the metastable well with a probability close to one.
The whole process strongly depends on the history of the sample, in particular, on the
direction of the field sweep.
If the field is increasing, then at the $n,k$-th resonance the Mn$_{12}$ molecules which cross
the barrier satisfy
\begin{equation}\label{DeplIntDel}
\Delta_{nk,v} < \Delta_{nk} \qquad {\rm but} \qquad \Delta_{n+1,k} < \Delta_{n+1,k,v},
\end{equation}
where $\Delta_{nk,v}$ is given by Eq.\ (\ref{Delnkv}).
The molecules with $\Delta_{n+1,k,v} < \Delta_{n+1,k}$ cannot cross the barrier at the $n$th
resonance because they have already left the metastable well at the $(n+1)$-th resonance.
The inequalities in Eq.\ (\ref{DeplIntDel}) determine the range of the logarithm of the
transverse anisotropy $x=\ln\tilde E$
\begin{equation}\label{DeplIntx}
x_{nk,v} < x  < x_{n+1,k,v}
\end{equation}
for molecules that are going to escape at the $n$th resonance.
The fraction of these molecules, which determines the magnetization step at the resonance, is
given by
\begin{equation}\label{pnkv}
P_{nk,v} = \int_{x_{nk,v}}^{x_{n+1,k,v}} dx f_L(x).
\end{equation}
If $H_z$ is increasing through a value at which the barrier disappears, all Mn$_{12}$
molecules will escape,
$\sum_{nk} P_{nk,v} = 1$.
With increasing temperature or lowering the sweep rate, the depletion effects become stronger
and stronger, which leads to the disappearance of tunneling resonances at certain $n$ and $k$.
Mathematically it manifests itself in the disappearance of the $x$-interval in Eq.\
(\ref{DeplIntx}), which leads to $P_{nk,v}=0$.

To see which resonances are active and which are not, it is convenient to plot $x_{nk,v}$ of
Eq.\ (\ref{xnkvDef}) versus $n$ for different $k$, Figs.\ \ref{fig_xv07} and \ref{fig_xv1}.
For $dH_z/dt >0$, the resonances are crossed in the order from small $k$ to large $k$ and,
within a given $k$-resonance, from the right to the left.
If the value of $x_{nk,v}$ decreases, the $x$-interval in Eq.\ (\ref{DeplIntx}) does exist and
the resonance is active.
In the opposite case the resonance vanishes.
This happens for the resonances with high $k$ in Figs.\ \ref{fig_xv07} and \ref{fig_xv1}.
Note that the maximal value of $n$ for each $k$-resonance, $n_{\rm max}$, is the greatest $n$
satisfying $\qquad \tilde v < \tilde v_{nk}^{\rm max}$.
In the limit $T\to 0$ the maximal velocity $v_{\rm max}$ of Eq.\ (\ref{vmaxdef}) goes to zero
for all $n>0$.
As a result, on lowering $T$, the branches of $x_{nk,v}$ in Figs.\ \ref{fig_xv07} and
\ref{fig_xv1} become shorter and shorter and eventually reduce to the set of $n=0$ points
along the vertical axis.
This is in accordance with the obvious fact that at $T=0$ only transitions from the metastable
ground state, $n=0$, are possible.
With increasing temperature, the branches of  move down for $n>0$, and the ground-state
resonances disappear one after the other.
The same occurs if the sweep rate $\tilde v$ decreases.
The dependence of $x_{nk,v}$ on $\tilde v$, however, is only logarithmic [see Eq.\
(\ref{xnkvDef})], so that the disappearance of the ground-state resonances at temperatures
well below 1K requires unrealistically small values of $\tilde v$.

\begin{figure}[t]
\unitlength1cm
\begin{picture}(11,6.5)
\centerline{\psfig{file=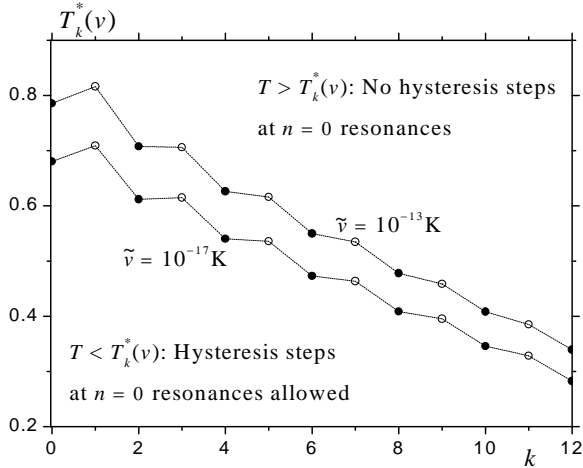,angle=-90,width=9cm}}
\end{picture}
\caption{ \label{fig_tstar} The values $T^*_k(\tilde v)$ separating regions with and without
hysteresis steps at $n=0$ resonances. }
\end{figure}

Note that the slope of the branches $x_{nk,v}$ in Figs.\ \ref{fig_xv07} and \ref{fig_xv1}
changes for a certain combination of parameters $k$, $T$, or $\tilde v$ for which the
dependence $x_{nk,v}$ on $n$ is nearly flat.
One can estimate where it happens from the condition $x_{0k,v}=x_{1k,v}$.
For the temperature $T^*_k(\tilde v)$ corresponding to this transition we get
\begin{eqnarray}\label{TStarDef}
&&
 T^*_k(\tilde v) \cong \frac{D(2S-1-k)} { \ln Q_v}
\nonumber\\
&&%
Q_v\equiv \frac{g_{1k}^2\delta\varepsilon_{0k}}{g_{0k}^2\delta\varepsilon_{1k}}
\left(\frac{\pi g_{0k}^2}{2\tilde v \delta\varepsilon_{0k}}\right)^{1/\xi_{0k}} .
\end{eqnarray}
This dependence is shown in Fig.\ \ref{fig_tstar} for two different sweep rates corresponding
to the boundaries of a typical experimental window.
In Mn$_{12}$ crystals with dislocations, temperature $ T^*_k(\tilde v)$ plays the role similar
to that of temperature $T_{00}$ for an ideal Mn$_{12}$ crystal [cf.\ Eq.\ (6.1) of Ref.\
\onlinecite{garchu97}].
The latter is the boundary between the pure ground-state tunneling and the thermally assisted
tunneling.
It can be obtained by equating $\Delta_{0k}^2/\delta\varepsilon_{0k}$ and
$(\Delta_{1k}^2/\delta\varepsilon_{1k}) e^{-\Delta\varepsilon_{1k}/T}$.
With the help of Eq.\ (\ref{SplEvenRewr}) one obtains
\begin{equation}\label{T00}
T_{00,k}=\frac {D(2S-1-k)} {\ln Q_E}, \qquad Q_E \equiv
\frac{g_{1k}^2\delta\varepsilon_{0k}}{g_{0k}^2\delta\varepsilon_{1k}} \frac 1 {\tilde E^2 }.
\end{equation}
In a crystal with dislocations the anisotropy $\tilde E$ is distributed and its role is played
by the sweep rate $\tilde v$ that ``chooses'' from the distribution of $\tilde E$ the matching
value of $\tilde E$  satisfying Eq.\ (\ref{DelnkDef}) with $n=0$.
Substitution of the solution of that equation,  Eq.\ (\ref{xnkvDef}), into Eq.\ (\ref{T00}),
gives Eq.\ (\ref{TStarDef}).

The temperature $T_{00,k}$ is close to the quantum-classical transition temperature $T_0$ (see
Refs.\ \onlinecite{garchu97,chugar97}) if the transition is first order.
Since for the model with transverse anisotropy the boundary between the first- and
second-order transitions is $E_0=D/3$ (Ref.\ \onlinecite{liamueparzim98}), which corresponds
to $\tilde E_0=1/6$, the values $\tilde E \ll 1$ in the main part of the Mn$_{12}$ crystal
fall into the range of first-order transition.
Accordingly, the change between the two regimes in Figs.\ \ref{fig_xv07} and \ref{fig_xv1} is
abrupt: The minimum of $x_{nk,v}$ at $n=0$ is replaced by the minimum at $n=n_{\rm max}$, at
$T\approx  T^*_k(\tilde v)$.
To study second-order transitions, one should use much greater sweep rates $\tilde v$ which
will shift the curves in Figs.\ \ref{fig_xv07} and \ref{fig_xv1} to higher values of
transverse anisotropies $\tilde E$.
This alone, however, does not guarantee that $P_{nk,v}$ of Eq.\ (\ref{pnkv}) is large enough
for the effect to be observed.
For realistic concentrations of dislocations the values of $f_L$ in that range of $\tilde E$
are very small, see Fig.\ \ref{fig_distribe}.

\section{Relaxation in the thermally activated regime}
\label{sec_activated}

As was commented below Eq.\ (\ref{fnkDefImproved}), at temperatures above the
quantum-classical transition, $T>T_0$ the spins of Mn$_{12}$ molecules escape over the top of
the barrier through $n=n_{\rm top}$, since this value of $n$ maximizes the function $f_{nk}$.
It is interesting to note, however, that Eq.\ (\ref{fnkDefImproved}) describes Landau-Zener
transitions in the field-sweep setup, thus Eqs.\ (\ref{NBeforeAfterRewr}) and
(\ref{NnBeforeAfter}) are only valid if it is the resonant tunneling that is dominating the
escape.
If the spins escape over the top of the barrier, it is no longer important whether the levels
are in resonance or not, so that the steps in the dynamic hysteresis curves due to the
Landau-Zener transitions should be washed out.

\begin{figure}[t]
\unitlength1cm
\begin{picture}(11,6.5)
\centerline{\psfig{file=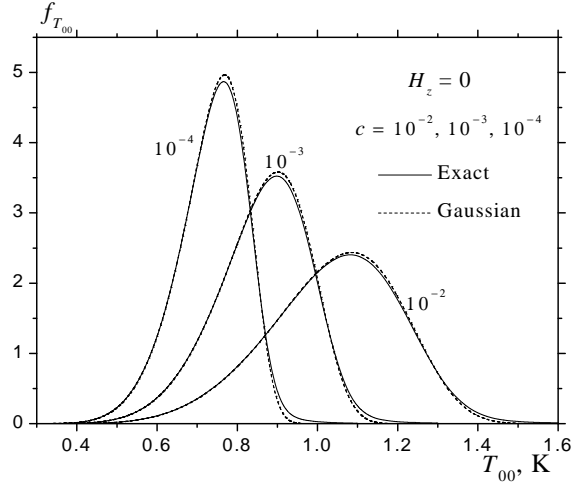,angle=-90,width=9cm}}
\end{picture}
\caption{ \label{fig_distt00} Distribution of the transition temperatures $T_{00}$ of Eq.\
(\protect\ref{T00}) in Mn$_{12}$ crystals with dislocations. Dashed lines correspond to the
Gaussian approximation of Eq.\ (\protect\ref{GaussianEL})}.
\end{figure}

Resonant transitions can still be detected at $T \gtrsim T_0$ due to the reduction of the
effective energy barrier in measurements of the time relaxation and of the linear dynamic
susceptibility because tunneling at resonance occurs via the level pair for which
$\Delta_{nk}$ is comparable with the level width $\Gamma_{nn'}$ (Ref.\ \onlinecite{garchu97}).
These levels are lower than those at the top of the barrier that satisfy $\Delta_{nk}\sim
\delta\varepsilon_{nk}$.
However, in the field-sweep setup the main contribution to the escape rate comes from the
levels at the top of the barrier, which makes the escape process non-resonant.
A seeming paradox is that resonant tunneling transitions are observed in the dynamic
hysteresis experiments in the activation regime.\cite{frisartejzio96}
This paradox can be resolved if one takes into account the distribution of $T_0$ in a
non-ideal Mn$_{12}$ crystal.
Then some of Mn$_{12}$ molecules have $T_0<T$, while others have  $T_0>T$, making tunneling
resonances weaker but still present as long as there are molecules with $T_0>T$.
Since for realistic concentrations of dislocations the typical values of the transverse
anisotropy satisfy $E\ll D$, the quantum-classical transition is first order and a good
estimation for $T_0$ is the temperature $T_{00}$ given by Eq.\ (\ref{T00}).
The distribution of $T_{00}$ can be obtained from the distribution of $\ln \tilde E$, see
Eqs.\ (\ref{fLDef}) and (\ref{GaussianEL}).
This distribution has an appreciable width, especially for high concentration of dislocations,
see Fig.\ \ref{fig_distt00}.
In experiments, quantum steps in the hysteresis have been observed at temperatures as high as
2.8K.
This may be either an indication that the fine tuning of the theory (see below) is needed or
an indication that Mn$_{12}$ crystals, or even molecules themselves, contain some stronger
defects than studied in this paper.
Note that for $E=E_0=D/3$ (which is the boundary between the first- and second-order
transition) one has $T_0=SD/\pi\approx 2$K (Ref.\ \onlinecite{liamueparzim98}).

Since the first-order transition is sharp, there are two groups of Mn$_{12}$ molecules:
Molecules of the first group escape via thermal activation over the top of the barrier, while
molecules of the second group escape via the ground-state tunneling.
Theoretical analysis of this situation is more cumbersome since the basic Eqs.\
(\ref{NBeforeAfterRewr}) and (\ref{NnBeforeAfter}) are valid only for a fraction of Mn$_{12}$
molecules in the crystal.
It is thus better to step back to Eq.\ (\ref{KinEq}) that yields a greater slope of the
dynamic hysteresis curve at resonances and smaller but nonzero slope off resonances.
We do not attempt to solve this problem in this article.
Instead, we will consider for simplicity the time relaxation off resonance in the activation
regime using the classical model, to demonstrate that the relaxation is non-exponential due to
dislocations.

For the classical model with transverse anisotropy $E\ll D$, the energy barrier is given by
\begin{eqnarray}\label{UClass}
&&
 U =U(E) \cong U_0 - \Delta U,
 \nonumber\\
 &&
 U_0 = DS^2(1-h_z)^2, \qquad  \Delta U = |E|S^2(1-h_z^2),
\end{eqnarray}
where $h_z=H_z/[2S(D-|E|)]$.
If transverse anisotropies are distributed the magnetization relaxation curve in the Arrhenius
regime is given by
\begin{equation}\label{RtDef}
R(t) = 2\int_0^\infty d E f_E(E) \exp[-\Gamma(E)t],
\end{equation}
where
\begin{equation}\label{GammaDef}
\Gamma(E) = \Gamma_0 e^{-U(E)/T}.
\end{equation}
The transverse-anisotropy distribution $f_E(E)$ has been calculated in Sec.\ \ref{sec_random}
for a random array of linear dislocations.
Using the results of that section, one can rewrite $R(t)$ in the scaling form
\begin{equation}\label{RtScaled}
R(\tilde t) = 2\int_0^\infty d \alpha f_\alpha(\alpha) \exp[-\tilde t e^{p\alpha}],
\end{equation}
where $\tilde t \equiv t/\tau$, $\tau^{-1}=\Gamma_0 e^{-U_0/T}$, and
\begin{equation}\label{pDef}
p\equiv \frac {1+h_z}{1-h_z} \frac {2 U_0 \tilde E_c} T .
\end{equation}
If $p\to 0$ in Eq.\ (\ref{RtScaled}), then one returns to the simple exponential relaxation,
$R(\tilde t) = \exp(-\tilde t)$.
In the case of $p\gg 1$ one has $R(\tilde t) \cong 2\int_0^{\alpha_t} d \alpha
f_\alpha(\alpha)$, where $\alpha_t = (1/p)\ln(1/\tilde t)$, i.e., relaxation is
logarithmically stretched.
For Mn$_{12}$ in the kelvin range, $U_0/T$ is large but the value of $\tilde E_c$ is small for
a realistic concentration of dislocations $c$, so that $p$ is typically of order unity.
Using estimates from the final part of Sec.\ \ref{sec_random}, at $H_z=0$ and $T=1$K one
obtains $p=1.94$ for $c=10^{-2}$, $p=0.61$ for $c=10^{-3}$, and $p=0.194$ for $c=10^{-4}$.
Thus Eq.\ (\ref{RtScaled}) does not simplify and it should be computed numerically with the
use of Eq.\ (\ref{falphaRes}).
Note that for large crystals $f_\alpha$ can be approximated by the Gaussian of Eq.\
(\ref{Gaussian}), which implies that the actual parameter of the problem is $\tilde p = 2\sqrt
L p\approx 6.5 p$.

The short-time behavior of $R(\tilde t)$ is singular due to Mn$_{12}$ molecules [see Eq.\
(\ref{n1})] which are close to dislocations.
These molecules correspond to the asymptote of the anisotropy-distribution function, $f_\alpha
\cong 1/\alpha^3$, for $\alpha \gtrsim \alpha_1$, which yields
\begin{equation}\label{RtLog}
R(\tilde t)\cong 1-\left[ \frac p {\ln(1/\tilde t)}\right]^2 .
\end{equation}
The singularity is rather weak for large crystals since the fraction $n_1$ of molecules which
relax according to Eq.\ (\ref{RtLog}) is small.
One can find the time $\tilde t_1$ at which this stage of relaxation is completed from the
condition $R(\tilde t_1) = 1-n_1$ that gives
\begin{equation}\label{t1}
\tilde t_1 = \exp[-p/\sqrt{n_1}] = \exp\left[-p\sqrt{4L\ln(4L/\sqrt{\pi})}\right].
\end{equation}
It is exponentially small for $p\gg 1$ and/or for large crystals.
If [see comment after Eq.\ (\ref{n1})] one replaces $\alpha_1$ by three $3\alpha_1$, the
exponent of Eq.\ (\ref{t1}) will increase by a factor of three.
This means that the actual value of $t_1$ is much smaller and the best way to find it is to
compare Eq.\ (\ref{RtLog}) with the exact relaxation curve of Eq.\ (\ref{RtScaled}).
Note, however, that in natural units this time is $t_1 =\tau\tilde t_1$, where $\tau$ is the
relaxation time of an ideal Mn$_{12}$ crystal which is exponentially long for $p\gg 1$.

\begin{figure}[t]
\unitlength1cm
\begin{picture}(11,6.5)
\centerline{\psfig{file=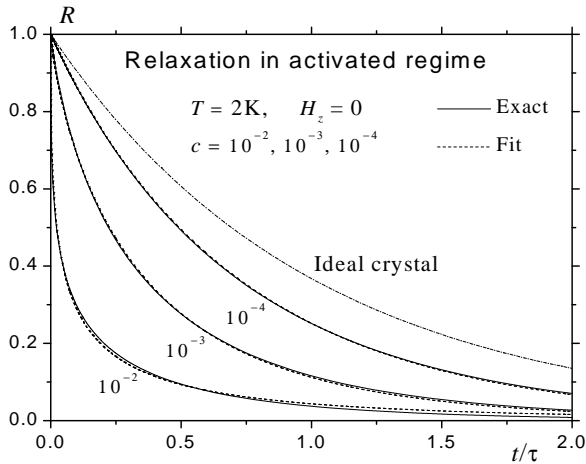,angle=-90,width=9cm}}
\end{picture}
\caption{ \label{fig_rt} Magnetization relaxation curves for Mn$_{12}$ with dislocations in
the activated regime.
The fitted values of $\zeta$ decrease on lowering temperature and increasing $c$. }
\end{figure}

Numerically computed relaxation curves $R(\tilde t)$ for $H_z=0$, $T=2$K, and different
concentrations of dislocations $c$ are shown in Fig.\ \ref{fig_rt}.
One can see that deviations of $R(\tilde t)$ from a pure exponential are quite pronounced.
For not too large concentrations $c$ one can fit  $R(\tilde t)$ by a stretched exponential
\begin{equation}\label{RtStretched}
R(\tilde t) = \exp(-a \tilde t^\zeta).
\end{equation}
Although this dependence does not follow from any theory, the fits are surprisingly good.
It would be interesting to see if the fit of the relaxation curve with Eq.\
(\ref{RtStretched}) provides the actual concentration of dislocations in a Mn$_{12}$ crystal
found by other methods.
For this purpose, we listed the fitted values of $a$ and $\zeta$, together with the parameter
$p$ of Eq.\ (\ref{pDef}), in Table \ref{tab_axi} for Mn$_{12}$ crystals of the typical size
$0.5\times 0.5$ mm$^2$ at $T=2$K and $H_z=0$.
For other temperatures and magnetic fields, one can interpolate $a$ and $\zeta$ on $p$ using
Table \ref{tab_axi}.
Note that in our picture the $\exp(-\sqrt{t})$ relaxation observed in some experiments is one
of many possibilities corresponding to different fields, temperatures, and concentrations.

\begin{table}
\caption{Parameter $p$ of Eq.\ (\protect\ref{pDef}) and fitting parameters $a$ and $\zeta$ for
$0.5\times 0.5$ mm$^2$ Mn$_{12}$ crystals at $T=2$K, $H_z=0$, and  different concentrations of
dislocations $c$. \label{tab_axi} }
\begin{tabular}{cccc}
 $c$ &$p$  &$a$ &$\zeta$\\
\tableline
$10^{-4}$& 0.097 & 1.38 & 0.97  \\
$3\times 10^{-4}$& 0.168 & 1.68 & 0.91  \\
$10^{-3}$& 0.306 & 2.19 & 0.77  \\
$3\times 10^{-3}$& 0.530 & 2.68 & 0.59  \\
$10^{-2}$& 0.97 & 3.13 & 0.40  \\
\end{tabular}
\end{table}

\section{Conclusions}
\label{sec_conclusion}

We have developed a comprehensive theory of quantum spin relaxation in Mn$_{12}$ acetate
crystals, which takes into account imperfections of the crystal structure and is based upon
the generalization of the Landau-Zener effect for incoherent tunneling from excited energy
levels.
All experimental features of the low-temperature magnetic behavior of Mn$_{12}$ crystals find
natural explanation within this theoretical framework.

Linear dislocations at plausible concentrations provide the transverse anisotropy which is the
main source of tunneling in Mn$_{12}$.
Local rotations of the easy axis due to dislocations result in a transverse magnetic field for
any external field applied along the $c$-axis of the crystal.
For odd resonances, the tunneling matrix element contains the transverse field only in the
first order of the perturbation theory, while the tranverse anisotropy makes the principal
contribution to the transition amplitude.
This explains the presence of odd tunneling resonances and their strength relative to that of
even resonances.

Transverse anisotropies and fields are distributed in the crystal.
One consequence of that distribution is that the temperature of the crossover between quantum
and thermal behavior becomes distributed within an appreciable range.
This may be in part responsible for the experimental fact that quantum steps in the hysteresis
of Mn$_{12}$ are well pronounced in the thermally activated regime.

Crystal defects produce a broad distribution of tunnel splittings that can be extracted from
the dependence of the magnetic relaxation on the field-sweep rate.
The theory predicts that relaxation curves for different tunneling resonances can be scaled
onto one master curve. \cite{myriampriv}
The first derivative of this curve equals the distribution function of transverse anisotropies
in the crystal.

Due to that distribution, quantum steps in the magnetization curve should appear and disappear
in a peculiar manner.
At a given temperature and field-sweep rate, our theory predicts which steps must be absent
\cite{myriampriv} regardless of the distribution function.

Another consequence of the distribution is that the magnetic relaxation in the thermally
activated regime follows the stretched-exponential law.
The exponent in this law depends on the field, temperature and concentration of defects.
In zero field at $T=2$K it is between 0.97 and 0.40 for concentrations ranging from $10^{-4}$
to $10^{-2}$ per unit cell of the crystal.

\section*{Acknowledgments}

We thank Myriam Sarachik and members of her experimental group for numerous illuminating
discussions and for providing us with unpublished experimental data on Mn$_{12}$.
This work has been supported by the NSF Grant No. 9978882.


\end{document}